\documentclass[12pt,a4paper]{article}
\pdfoutput=1

\usepackage{ifthen} 
\newboolean{pdflatex}
\setboolean{pdflatex}{true} 

\newboolean{articletitles}
\setboolean{articletitles}{true} 

\newboolean{uprightparticles}
\setboolean{uprightparticles}{false} 

\newboolean{inbibliography}
\setboolean{inbibliography}{false} 

\newboolean{paperconf}
\setboolean{paperconf}{true} 

\ifthenelse{\boolean{paperconf}}{
\def\paperauthors{LHCb collaboration}
}{
\def\paperauthors{A.~N.~Other}
}
\def\paperasciititle{A measurement of the CP asymmetry difference between Lambda_c+ to ph-h+ decays} 
\def\papertitle{A measurement of the $C\!P$ asymmetry difference between $\varLambda_{c}^{+} \to pK^{-}K^{+}$ and $p\pi^{-}\pi^{+}$ decays} 
\def\paperkeywords{{High Energy Physics}, {LHCb}, {Charm}, {CP violation}} 
\def\papercopyright{CERN on behalf of the LHCb collaboration}
\def\paperlicence{CC-BY-4.0}
\def\paperlicenceurl{https://creativecommons.org/licenses/by/4.0/}

\usepackage[top=1in, bottom=1.25in, left=1in, right=1in]{geometry}

\usepackage{microtype}
\usepackage{lineno}  
\usepackage{xspace} 
\usepackage{caption}

\usepackage{graphicx}  
\usepackage{color}
\usepackage{colortbl}
\usepackage{booktabs}
\usepackage{adjustbox}
\graphicspath{{./figures/}} 
\usepackage{tikz}
\usepackage{tikz-3dplot}

\usepackage{amsmath} 
\usepackage{amssymb}
\usepackage{amsfonts}
\usepackage{upgreek} 

\newcommand*\patchAmsMathEnvironmentForLineno[1]{
\expandafter\let\csname old#1\expandafter\endcsname\csname #1\endcsname
\expandafter\let\csname oldend#1\expandafter\endcsname\csname
end#1\endcsname
 \renewenvironment{#1}
   {\linenomath\csname old#1\endcsname}
   {\csname oldend#1\endcsname\endlinenomath}
}
\newcommand*\patchBothAmsMathEnvironmentsForLineno[1]{
  \patchAmsMathEnvironmentForLineno{#1}
  \patchAmsMathEnvironmentForLineno{#1*}
}
\AtBeginDocument{
\patchBothAmsMathEnvironmentsForLineno{equation}
\patchBothAmsMathEnvironmentsForLineno{align}
\patchBothAmsMathEnvironmentsForLineno{flalign}
\patchBothAmsMathEnvironmentsForLineno{alignat}
\patchBothAmsMathEnvironmentsForLineno{gather}
\patchBothAmsMathEnvironmentsForLineno{multline}
\patchBothAmsMathEnvironmentsForLineno{eqnarray}
}

\usepackage{siunitx}
\usepackage{hyperxmp}

\usepackage[pdftex]{hyperref}
\usepackage[all]{hypcap}

\usepackage{xspace} 
\usepackage{upgreek}

\def\lhcb {\mbox{LHCb}\xspace}

\def\lhc    {\mbox{LHC}\xspace}

\def\MagUp {\mbox{\em Mag\kern -0.05em Up}\xspace}

\ifthenelse{\boolean{uprightparticles}}
{

 \def\Pmu         {\ensuremath{\upmu}\xspace}

 \def\Ppi         {\ensuremath{\uppi}\xspace}

 \def\PDelta      {\ensuremath{\Delta}\xspace}                 
 \def\PXi      {\ensuremath{\Xi}\xspace}                 
 \def\PLambda      {\ensuremath{\Lambda}\xspace}                 
 \def\PSigma      {\ensuremath{\Sigma}\xspace}                 
 \def\POmega      {\ensuremath{\Omega}\xspace}                 
 \def\PUpsilon      {\ensuremath{\Upsilon}\xspace}

 \def\PB      {\ensuremath{\mathrm{B}}\xspace}                 
                  
 \def\PD      {\ensuremath{\mathrm{D}}\xspace}

 \def\PK      {\ensuremath{\mathrm{K}}\xspace}

 \def\Pb      {\ensuremath{\mathrm{b}}\xspace}                 
 \def\Pc      {\ensuremath{\mathrm{c}}\xspace}

 \def\Pi      {\ensuremath{\mathrm{i}}\xspace}

 \def\Pp      {\ensuremath{\mathrm{p}}\xspace}

 \def\Ps      {\ensuremath{\mathrm{s}}\xspace}

}
{

 \def\Pmu         {\ensuremath{\mu}\xspace}

 \def\Ppi         {\ensuremath{\pi}\xspace}

 \mathchardef\PDelta="7101
 \mathchardef\PXi="7104
 \mathchardef\PLambda="7103
 \mathchardef\PSigma="7106
 \mathchardef\POmega="710A
 \mathchardef\PUpsilon="7107
                  
 \def\PB      {\ensuremath{B}\xspace}                 
                  
 \def\PD      {\ensuremath{D}\xspace}

 \def\PK      {\ensuremath{K}\xspace}

 \def\Pb      {\ensuremath{b}\xspace}                 
 \def\Pc      {\ensuremath{c}\xspace}

 \def\Pi      {\ensuremath{i}\xspace}

 \def\Pp      {\ensuremath{p}\xspace}

 \def\Ps      {\ensuremath{s}\xspace}

}

\makeatletter
\ifcase \@ptsize \relax
  \newcommand{\miniscule}{\@setfontsize\miniscule{4}{5}}
\or
  \newcommand{\miniscule}{\@setfontsize\miniscule{5}{6}}
\or
  \newcommand{\miniscule}{\@setfontsize\miniscule{5}{6}}
\fi
\makeatother

\DeclareRobustCommand{\optbar}[1]{\shortstack{{\miniscule (\rule[.5ex]{1.25em}{.18mm})}
  \\ [-.7ex] $#1$}}

\def\muon       {{\ensuremath{\Pmu}}\xspace}
\def\mup        {{\ensuremath{\Pmu^+}}\xspace}
\def\mun        {{\ensuremath{\Pmu^-}}\xspace}

\def\squark    {{\ensuremath{\Ps}}\xspace}

\def\cquark    {{\ensuremath{\Pc}}\xspace}

\def\bquark    {{\ensuremath{\Pb}}\xspace}

\def\pion   {{\ensuremath{\Ppi}}\xspace}

\def\pip    {{\ensuremath{\pion^+}}\xspace}
\def\pim    {{\ensuremath{\pion^-}}\xspace}

\def\kaon    {{\ensuremath{\PK}}\xspace}
  \def\Kbar    {{\kern 0.2em\overline{\kern -0.2em \PK}{}}\xspace}

\def\KorKbar    {\kern 0.18em\optbar{\kern -0.18em K}{}\xspace}

\def\Kzb     {{\ensuremath{\Kbar{}^0}}\xspace}
\def\Kp      {{\ensuremath{\kaon^+}}\xspace}
\def\Km      {{\ensuremath{\kaon^-}}\xspace}

\def\KS      {{\ensuremath{\kaon^0_{\mathrm{ \scriptscriptstyle S}}}}\xspace}

  \def\Dbar    {{\kern 0.2em\overline{\kern -0.2em \PD}{}}\xspace}
\def\D       {{\ensuremath{\PD}}\xspace}

\def\DorDbar    {\kern 0.18em\optbar{\kern -0.18em D}{}\xspace}
\def\Dz      {{\ensuremath{\D^0}}\xspace}

\def\Dp      {{\ensuremath{\D^+}}\xspace}

\def\Dsp     {{\ensuremath{\D^+_\squark}}\xspace}

\def\Bbar    {{\ensuremath{\kern 0.18em\overline{\kern -0.18em \PB}{}}}\xspace}

\def\BorBbar    {\kern 0.18em\optbar{\kern -0.18em B}{}\xspace}

  \def\Y#1S{\ensuremath{\PUpsilon{(#1S)}}\xspace}

\def\proton      {{\ensuremath{\Pp}}\xspace}

\def\Lz          {{\ensuremath{\PLambda}}\xspace}
\def\Lbar        {{\ensuremath{\kern 0.1em\overline{\kern -0.1em\PLambda}}}\xspace}
\def\LorLbar    {\kern 0.18em\optbar{\kern -0.18em \PLambda}{}\xspace}

\def\Lb      {{\ensuremath{\Lz^0_\bquark}}\xspace}

\newcommand{\decay}[2]{\mbox{\ensuremath{#1\!\to #2}}\xspace}         

\def\to                 {\ensuremath{\rightarrow}\xspace}

\def\CP                {{\ensuremath{C\!P}}\xspace}

\def\AT#1     {\ensuremath{A_{\mathrm{T}}^{#1}}\xspace}

\def\C#1      {\ensuremath{\mathcal{C}_{#1}}\xspace}                       
\def\Cp#1     {\ensuremath{\mathcal{C}_{#1}^{'}}\xspace}                    
\def\Ceff#1   {\ensuremath{\mathcal{C}_{#1}^{\mathrm{(eff)}}}\xspace}        
\def\Cpeff#1  {\ensuremath{\mathcal{C}_{#1}^{'\mathrm{(eff)}}}\xspace}       
\def\Ope#1    {\ensuremath{\mathcal{O}_{#1}}\xspace}                       
\def\Opep#1   {\ensuremath{\mathcal{O}_{#1}^{'}}\xspace}

\DeclareSIUnit\clight{\text{\ensuremath{c}}}
\DeclareSIUnit\micron{\micro\metre}
\DeclareSIUnit\mrad{\milli\radian}
\DeclareSIUnit\gauss{G}

\DeclareSIUnit\meV{\milli\eV}
\DeclareSIUnit\keV{\kilo\eV}
\DeclareSIUnit\MeV{\mega\eV}
\DeclareSIUnit\GeV{\giga\eV}
\DeclareSIUnit\TeV{\tera\eV}

\DeclareSIUnit[per-mode=symbol]\MeVc{\MeV\!\per\clight}
\DeclareSIUnit[per-mode=symbol]\GeVc{\GeV\!\per\clight}

\DeclareSIUnit[per-mode=symbol]\MeVcc{\MeV\!\per\clight\squared}
\DeclareSIUnit[per-mode=symbol]\GeVcc{\GeV\!\per\clight\squared}
\DeclareSIUnit[per-mode=symbol]\GeVGeVcccc{\GeV\squared\!\per\clight^{4}}

\DeclareSIUnit\mb{\micro\barn}
\DeclareSIUnit\nb{\nano\barn}
\DeclareSIUnit\pb{\pico\barn}
\DeclareSIUnit\fb{\femto\barn}
\DeclareSIUnit\ab{\atto\barn}
\DeclareSIUnit\zb{\zepto\barn}
\DeclareSIUnit\yb{\yocto\barn}

\DeclareSIUnit\invnb{\per\nano\barn}
\DeclareSIUnit\invpb{\per\pico\barn}
\DeclareSIUnit\invfb{\per\femto\barn}
\DeclareSIUnit\invab{\per\atto\barn}

\DeclareSIUnit\Xrad{\text{\ensuremath{X_{0}}}}
\DeclareSIUnit\NIL{\text{\ensuremath{\lambda_{\text{int}}}}}
\DeclareSIUnit\mip{MIP}

\newcommand{\chisq}{\ensuremath{\chi^2}\xspace}

\def\gsim{{~\raise.15em\hbox{$>$}\kern-.85em
          \lower.35em\hbox{$\sim$}~}\xspace}
\def\lsim{{~\raise.15em\hbox{$<$}\kern-.85em
          \lower.35em\hbox{$\sim$}~}\xspace}

\def\sPlot{\mbox{\em sPlot}\xspace}

\def\pt         {\mbox{$p_{\mathrm{ T}}$}\xspace}

\def\msq        {\ensuremath{m^2}\xspace}

\def\evtgen     {\mbox{\textsc{EvtGen}}\xspace}

\def\gauss      {\mbox{\textsc{Gauss}}\xspace}
\def\geant      {\mbox{\textsc{Geant4}}\xspace}

\def\photos     {\mbox{\textsc{Photos}}\xspace}

\def\pythia     {\mbox{\textsc{Pythia}}\xspace}

\def\tell1  {TELL1\xspace}
\def\ukl1   {UKL1\xspace}

 \newcommand{\hp}{\ensuremath{h^{+}}\xspace}
\newcommand{\hm}{\ensuremath{h^{-}}\xspace}
\newcommand{\Lcp}{\ensuremath{\Lz_{\cquark}^{+}}\xspace}
\newcommand{\Lcm}{\ensuremath{\Lbar_{\cquark}^{-}}\xspace}
\newcommand{\Lcpm}{\ensuremath{\Lz_{\cquark}^{\pm}}\xspace}

\newcommand{\KK}{\ensuremath{\Km\Kp}\xspace}
\newcommand{\pipi}{\ensuremath{\pim\pip}\xspace}
\newcommand{\Kpi}{\ensuremath{\Km\pip}\xspace}
\newcommand{\hh}{\ensuremath{\hm\hp}\xspace}
\newcommand{\pKK}{\ensuremath{p\KK}\xspace}
\newcommand{\ppipi}{\ensuremath{p\pipi}\xspace}
\newcommand{\pKpi}{\ensuremath{p\Kpi}\xspace}
\newcommand{\phh}{\ensuremath{p\hh}\xspace}
\newcommand{\Lpi}{\ensuremath{\Lz\pip}\xspace}

\newcommand{\LcTopKK}{\ensuremath{\decay{\Lcp}{\pKK}}\xspace}
\newcommand{\LcToppipi}{\ensuremath{\decay{\Lcp}{\ppipi}}\xspace}
\newcommand{\LcTophh}{\ensuremath{\decay{\Lcp}{\phh}}\xspace}
\newcommand{\LbToLcmuX}{\ensuremath{\decay{\Lb}{\Lcp\mun{}X}}\xspace}
\newcommand{\DpToKpipi}{\ensuremath{\decay{\Dp}{\Km\pip\pip}}\xspace}
\newcommand{\DpToKKpi}{\ensuremath{\decay{\Dp}{\Km\Kp\pip}}\xspace}
\newcommand{\DsToKpipi}{\ensuremath{\decay{\Dsp}{\pim\Kp\pip}}\xspace}
\newcommand{\DsToKKpi}{\ensuremath{\decay{\Dsp}{\Km\Kp\pip}}\xspace}
\newcommand{\LcTopKz}{\ensuremath{\decay{\Lcp}{\proton\Kzb}}\xspace}
\newcommand{\LcToLpi}{\ensuremath{\decay{\Lcp}{\Lpi}}\xspace}
\newcommand{\LcTopKpi}{\ensuremath{\decay{\Lcp}{\pKpi}}\xspace}
\newcommand{\phiToKK}{\ensuremath{\decay{\phi}{\KK}}\xspace}

\newcommand{\ACP}{\ensuremath{A_{\CP}}\xspace}
\newcommand{\ACPwgt}{\ensuremath{A_{\CP}^{\text{wgt}}}\xspace}
\newcommand{\dACP}{\ensuremath{\Delta\ACP}\xspace}
\newcommand{\dACPwgt}{\ensuremath{\Delta\ACPwgt}\xspace}
\newcommand{\ARaw}{\ensuremath{A_{\text{raw}}}\xspace}
\newcommand{\ARawwgt}{\ensuremath{A_{\text{raw}}^{\text{wgt}}}\xspace}
\newcommand{\ARawBkg}{\ensuremath{A_{\text{raw}}^{\text{Bkg}}}\xspace}
\newcommand{\AP}{\ensuremath{A_{\text{P}}}\xspace}
\newcommand{\APLb}{\ensuremath{\AP^{\Lb}}\xspace}
\newcommand{\AD}{\ensuremath{A_{\text{D}}}\xspace}
\newcommand{\ADmu}{\ensuremath{\AD^{\muon}}\xspace}
\newcommand{\ADf}{\ensuremath{\AD^{f}}\xspace}

\newcommand{\msqtwo}[2]{\ensuremath{\msq(#1#2)}}
\newcommand{\msqphm}{\msqtwo{\proton}{\hm}\xspace}

\newcommand{\msqhh}{\msqtwo{\hm}{\hp}\xspace}

\newcommand{\sqrts}{\ensuremath{\sqrt{s}}\xspace}
\newcommand{\sqrtseq}[1]{\ensuremath{\sqrt{s} = \SI{#1}{\TeV}}\xspace}
\newcommand{\pp}{\ensuremath{\proton\proton}\xspace}

\newcommand{\eff}{\ensuremath{\varepsilon}\xspace}

\newcommand{\fitvar}{\ensuremath{m}\xspace}
\newcommand{\ftot}{\ensuremath{f_{\text{Tot}}}\xspace}

\newcommand{\paramvec}{\ensuremath{\xi}\xspace}
\newcommand{\Nsig}{\ensuremath{N_{\text{Sig}}}\xspace}
\newcommand{\Nbkg}{\ensuremath{N_{\text{Bkg}}}\xspace}

\newcommand{\bini}{\ensuremath{W_{i}}\xspace}
\newcommand{\binierr}{\ensuremath{\delta{}\bini}\xspace}

\newcommand{\paramvecp}{\ensuremath{\xi^{+}}\xspace}
\newcommand{\paramvecm}{\ensuremath{\xi^{-}}\xspace}
\newcommand{\Nsigpm}{\ensuremath{\Nsig^{\pm}}\xspace}
\newcommand{\Nbkgpm}{\ensuremath{\Nbkg^{\pm}}\xspace}

\newcommand{\binipm}{\ensuremath{\bini^{\pm}}\xspace}

\usepackage[nameinlink,capitalise]{cleveref}

\usepackage{cite} 
\usepackage{mciteplus}
 \usepackage{longtable}

\def\deltaACPval {0.30}
\def\deltaACPunc {0.91}
\def\deltaACPunit {\percent}

\def\ARawpKKval {3.72}
\def\ARawpKKunc {0.78}
\def\ARawpKKunit {\percent}
\def\ARawpKK {\SI{\ARawpKKval \pm \ARawpKKunc}{\ARawpKKunit}}
\def\ARawppipival {3.42}
\def\ARawppipiunc {0.47}
\def\ARawppipiunit {\percent}
\def\ARawppipi {\SI{\ARawppipival \pm \ARawppipiunc}{\ARawppipiunit}}

\hypersetup{
    pdfauthor={\paperauthors},
    pdftitle={\paperasciititle},
    pdfkeywords={\paperkeywords},
    pdfcopyright={Copyright (C) \papercopyright},
    pdflicenseurl={\paperlicenceurl}
}

\begin{document}

\renewcommand{\thefootnote}{\fnsymbol{footnote}}
\setcounter{footnote}{1}

\begin{titlepage}
\pagenumbering{roman}

\vspace*{-1.5cm}
\centerline{\large EUROPEAN ORGANIZATION FOR NUCLEAR RESEARCH (CERN)}
\vspace*{1.5cm}
\noindent
\begin{tabular*}{\linewidth}{lc@{\extracolsep{\fill}}r@{\extracolsep{0pt}}}
\ifthenelse{\boolean{pdflatex}}
{\vspace*{-1.5cm}\mbox{\!\!\!\includegraphics[width=.14\textwidth]{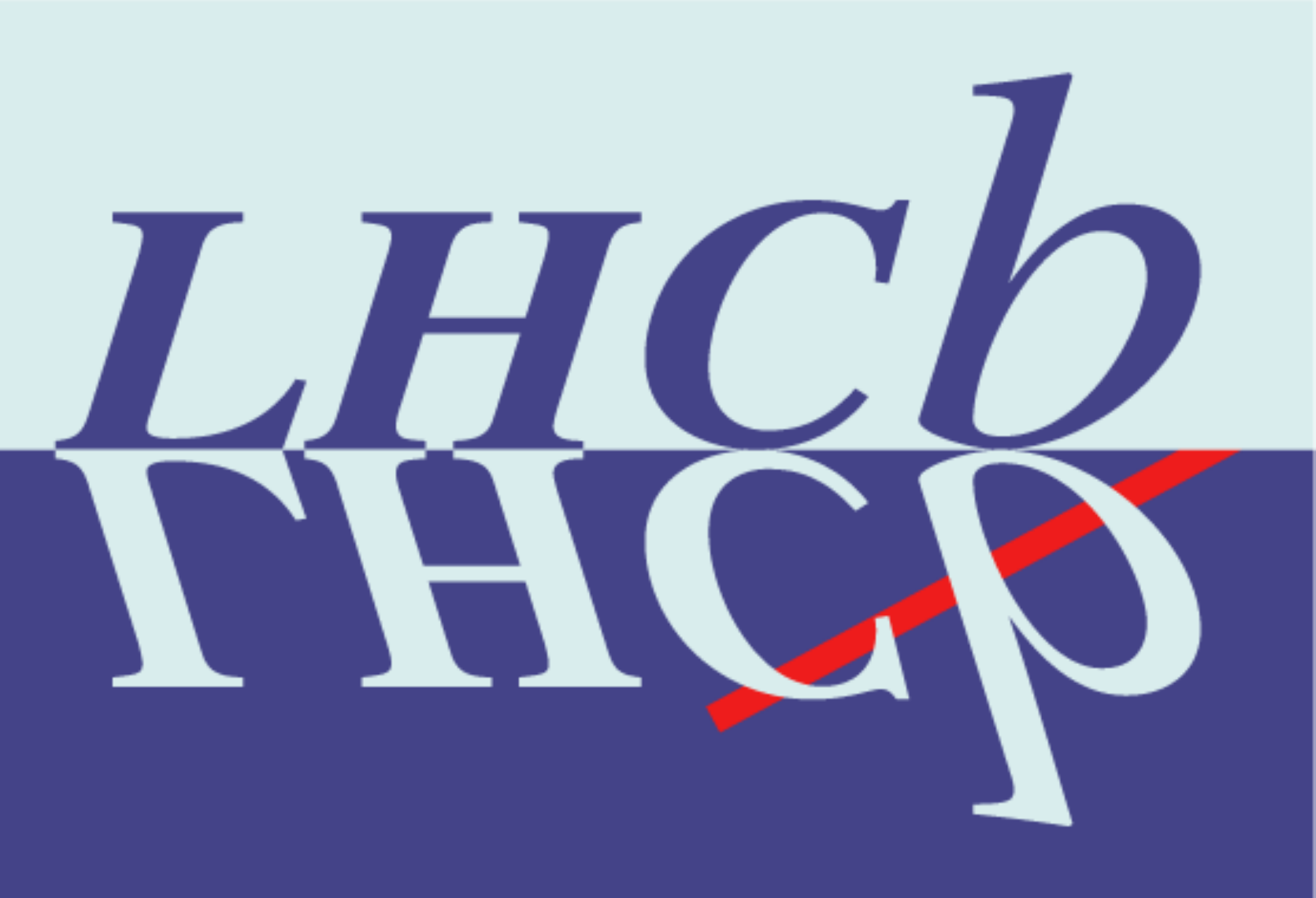}} & &}
{\vspace*{-1.2cm}\mbox{\!\!\!\includegraphics[width=.12\textwidth]{lhcb-logo.eps}} & &}
\\
 & & CERN-EP-2017-316 \\  
 & & LHCb-PAPER-2017-044 \\  
 & & \today \\ 
 & & \\
\end{tabular*}

\vspace*{2.0cm}

{\normalfont\bfseries\boldmath\huge
\begin{center}
  \papertitle 
\end{center}
}

\vspace*{2.0cm}

\begin{center}
\paperauthors\footnote{Authors are listed at the end of this paper.}
\end{center}

\vspace{\fill}

\begin{abstract}
  \noindent
  The difference between the $C\!P$ asymmetries in the decays 
  \mbox{$\varLambda_{c}^{+} \to pK^{-}K^{+}$} and \mbox{$\varLambda_{c}^{+} \to p\pi^{-}\pi^{+}$} 
  is presented.
  Proton-proton collision data taken at centre-of-mass energies of $7$ and 
  $8\,\mathrm{TeV}$ collected by the LHCb detector in 2011 and 2012 are used, 
  corresponding to an integrated luminosity of $3\,\mathrm{fb}^{-1}$.
  The $\varLambda_{c}^{+}$ candidates are reconstructed as part of the 
  \mbox{$\varLambda_{b}^{0} \to \varLambda_{c}^{+}\mu^{-}X$} decay chain.
  In order to maximize the cancellation of production and detection asymmetries 
  in the difference, the final-state kinematic distributions of the two samples 
  are aligned by  applying phase-space-dependent weights  to the 
  \mbox{$\varLambda_{c}^{+} \to p\pi^{-}\pi^{+}$} sample.
  This alters the definition of the integrated $C\!P$ asymmetry to 
  $A_{C\!P}^{\text{wgt}}(p\pi^{-}\pi^{+})$.
  Both samples are corrected for reconstruction and selection efficiencies 
  across the five-dimensional $\varLambda_{c}^{+}$ decay phase space.
  The difference in $C\!P$ asymmetries is found to be
  \begin{align*}
    \Delta{A^{\text{wgt}}_{C\!P}} &= A_{C\!P}(pK^{-}K^{+}) - A_{C\!P}^{\text{wgt}}(p\pi^{-}\pi^{+})\\
                                  &= (\deltaACPval \pm \deltaACPunc \pm 0.61)\,\si{\deltaACPunit},
  \end{align*}
  where the first uncertainty is statistical and the second is systematic.
\end{abstract}

\vspace*{2.0cm}

\begin{center}
  Published in JHEP 03 (2018) 182
\end{center}

\vspace{\fill}

{\footnotesize 
\centerline{\copyright~\papercopyright, licence \href{\paperlicenceurl}{\paperlicence}.}}
\vspace*{2mm}

\end{titlepage}

\newpage
\setcounter{page}{2}
\mbox{~}

\cleardoublepage

\renewcommand{\thefootnote}{\arabic{footnote}}
\setcounter{footnote}{0}

\pagestyle{plain} 
\setcounter{page}{1}
\pagenumbering{arabic}

\section{Introduction}
\label{sec:intro}

The Standard Model~(SM) does not provide a source of charge-parity (\CP) 
symmetry violation large enough to explain the matter-antimatter asymmetry 
observed in the universe~\cite{Gavela:1993ts}.
The ongoing experimental effort in searching for \CP violation in particle 
decays aims to find effects that are not expected in the SM, such that new 
dynamics are required.
Whilst the existence of \CP violation in kaon and beauty meson decays is 
well established~\cite{PDG2017,HFLAV16}, no observation has been made in the 
analyses of beauty baryons or charm hadrons, although evidence of \CP violation 
has recently been claimed for the former~\cite{LHCb-PAPER-2016-030}.
The most precise searches for \CP violation in the charm sector have been made 
using self-conjugate, singly Cabibbo-suppressed~(SCS) decays of the neutral \Dz 
meson to \KK and \pipi final 
states~\cite{LHCb-PAPER-2014-013,LHCb-PAPER-2015-055}.
Such SCS decays can include significant contributions from loop-level 
amplitudes, within which new dynamics can enter.

This article reports a search for \CP violation in the decays of the \Lcp charm 
baryon to the SCS \pKK and \ppipi final states (generically referred to as 
\phh).\footnotemark
\footnotetext{
  Charge conjugation is implied throughout this article, except in the 
  definition of asymmetry terms.
}
The difference in \CP asymmetry between the two decays, \dACP, is measured in a 
manner similar to previous measurements using \Dz 
decays~\cite{LHCb-PAPER-2014-013,LHCb-PAPER-2015-055}.
There is little theoretical understanding of the dynamics of \LcTophh 
decays~\cite{Bigi:2012ev}, partly due to the unknown resonant structure of the 
five-dimensional~(5D) phase space and partly due to the historical lack of 
large experimental datasets, and so no predictions for the magnitude of \CP 
violation in \LcTophh decays are currently available.
As \CP violation may be dependent on the position in phase space, leading to 
locally significant effects, a multidimensional analysis would be required to 
be maximally sensitive to such behaviour, requiring assumptions on the as-yet 
unknown amplitude model and \Lcp polarisation.
The work presented here instead integrates over the phase space as a search for 
global \CP-violating effects.

The presented analysis uses proton-proton collision data taken at 
centre-of-mass energies of \sqrtseq{7} and \SI{8}{\TeV}, collected by the \lhcb 
experiment at the Large Hadron Collider~(\lhc) in 2011 and 2012, corresponding 
to an integrated luminosity of \SI{3}{\per\femto\barn}.
To reduce the level of backgrounds, candidate \LcTophh decays are reconstructed 
as part of the \LbToLcmuX decay chain, where $X$ represents any number of 
additional, unreconstructed particles.
The long lifetime of the \Lb baryon~\cite{PDG2017}, in comparison with that of 
the \Lcp baryon, allows for the suppression of backgrounds through the 
requirement of a $\Lcp\mun$ vertex that is displaced with respect to the 
primary \pp interaction.
The total dataset contains of the order of \num{e4} and \num{e5} reconstructed 
\pKK and \ppipi signal candidates, respectively.

The observed charge asymmetry $\ARaw(f)$ for each \Lcp final state $f$, 
reconstructed in association with a muon, is measured as the difference in \Lcp 
and \Lcm signal yields divided by their sum.
The quantity \ARaw includes contributions from the \CP asymmetry in the \Lcp 
decay, as well as asymmetries due to experimental effects such as the \Lb 
production asymmetry and the muon and hadron detection asymmetries.
These effects have been measured at 
LHCb~\cite{LHCb-PAPER-2015-032,LHCb-PAPER-2016-013,LHCb-PAPER-2014-013,LHCb-PAPER-2012-009}, 
but with large uncertainties.
Using them directly, to correct for the experimental asymmetries in \ARaw, 
would then result in large systematic uncertainties on the correction factors.
Instead, assuming that the asymmetries are, or can be made to be, mode 
independent, the difference $\dACP = \ARaw(\pKK) - \ARaw(\ppipi)$ is equal to 
the difference in the \Lcp decay asymmetries, as all other asymmetries cancel.
A weighting technique is used to equalise the $\pKK$ and $\ppipi$ sample 
kinematics, thereby improving the level of cancellation of the various 
production, reconstruction, and selection asymmetries in \dACP, the formalism 
for which is presented in \cref{sec:formalism}.
A description of the \lhcb detector and the analysis dataset is given in 
\cref{sec:detdata}, followed by a description of the statistical models used to 
determine the signal yields from the data in \cref{sec:mass_fit}.
The weighting method used for correcting the measurement for experimental 
asymmetries and the evaluation of the efficiency variation across the 5D \phh 
phase space are presented in \cref{sec:corrections}.
Systematic effects are considered and quantified in \cref{sec:systematics}.
The results of the analysis are given in \cref{sec:results}, and finally a 
summary is made in \cref{sec:summary}.
 
\section{Formalism}
\label{sec:formalism}

The \CP asymmetry in the decays of the \Lcp baryon to a given final state $f$ 
is
\begin{equation}
  \ACP(f) = \frac{
    \Gamma(f) - \Gamma(\bar{f})
  }{
  \Gamma(f) + \Gamma(\bar{f})
  },
\end{equation}
where $\Gamma(f)$ is the decay rate of the $\decay{\Lcp}{f}$ process, and 
$\Gamma(\bar{f})$ is the decay rate of the charge conjugate decay 
$\decay{\Lcm}{\bar{f}}$.
Rather than measure the individual decay rates, it is simpler to count the 
number of reconstructed decays, and so the asymmetry in the yields is defined 
as
\begin{equation}
  \ARaw(f) = \frac{
    N(f\mun) - N(\bar{f}\mup)
    }{
    N(f\mun) + N(\bar{f}\mup)
  },
  \label{eqn:formalism:araw}
\end{equation}
where $N$ is the number of signal candidates reconstructed in association with 
a muon.
This is labelled as the raw asymmetry of the decay because the measurement of 
the physics observable of interest, \ACP, is contaminated by several 
experimental asymmetries.
Assuming that each contributing factor is small, the raw asymmetry can be 
expressed to first order as the sum
\begin{equation}
  \ARaw(f) = \ACP(f) + \APLb(f\muon) + \ADmu(\muon) + \ADf(f),
  \label{eqn:formalism:araw_sum}
\end{equation}
where \APLb, \ADmu, and \ADf are the \Lb production asymmetry, muon detection 
asymmetry, and \Lcp final-state detection asymmetry, respectively.
A nonzero \Lb production asymmetry may arise for several reasons, such as the 
relative abundance of matter quarks in the \pp collision region.
A dependence on the reconstructed \Lb final state is introduced by the detector 
acceptance and the reconstruction and selection applied to that state, which 
alters the observed \Lb production phase space.
The two detection asymmetries may be nonzero due to the different interaction 
cross-sections of the matter and antimatter states with the \lhcb detector.
There may also be charge-dependent reconstruction and selection effects.
In all cases, an experimental asymmetry is assumed to be fully parameterised by 
the kinematics of the objects involved.
The asymmetry of interest, \ACP, is assumed to be dependent on $f$ but 
independent of \Lcp kinematics.
This motivates a measurement of the difference between raw asymmetries of two 
distinct \Lcp decay modes, chosen to be \pKK and \ppipi in this analysis,
\begin{align}
  \dACP &= \ACP(\pKK) - \ACP(\ppipi)\\
        &\approx \ARaw(\pKK) - \ARaw(\ppipi),
\end{align}
where the approximation tends to an equality as the kinematics between the 
final states become indistinguishable.

The observed kinematics of the $\pKK\mun$ and $\ppipi\mun$ final states are not 
expected to be equal given the different energy release and resonant structure 
of the two \Lcp decays.
To ensure similarity, the kinematic spectra of one state can be matched to that 
of another.
As around six times as many \ppipi signal candidates are found than \pKK signal 
candidates, as described in \cref{sec:mass_fit}, the $\ppipi\mun$ data are 
weighted to match the $\pKK\mun$ data, given that the statistical uncertainty 
on the $\ARaw(\pKK)$ measurement will be the dominant contribution to that on 
\dACPwgt.
Details of the weighting procedure are given in \cref{sec:corrections}.
The kinematic weighting may alter the physics asymmetry, as the \ppipi phase 
space can be distorted, and so it is a weighted asymmetry, \ACPwgt, that enters 
the measurement
\begin{align}
  \dACPwgt &= \ACP(\pKK) - \ACPwgt(\ppipi)\\
           &\approx \ARaw(\pKK) - \ARawwgt(\ppipi).
\end{align}
To allow for comparisons with theoretical models, a weighting function is 
provided in the supplementary material of this article which provides a weight 
for a given coordinate in the five-dimensional \LcToppipi\ phase space and 
mimics the transformation imposed by the kinematic weighting applied here.
The five dimensions are defined similarly to those in 
Ref.~\cite{Aitala:1999uq}, with the only difference being that the `beam axis' 
is replaced by the displacement vector pointing from the \pp collision vertex, 
the primary vertex~(PV), to the $\Lcp\mun$ vertex.
 
\section{Detector and dataset}
\label{sec:detdata}

The \lhcb detector~\cite{Alves:2008zz,LHCb-DP-2014-002} is a single-arm forward 
spectrometer covering the \mbox{pseudorapidity} range $2 < \eta <5$, designed 
for the study of particles containing \bquark or \cquark quarks.
The detector includes a high-precision tracking system consisting of a 
silicon-strip vertex detector surrounding the \pp interaction region, a 
large-area silicon-strip detector located upstream of a dipole magnet with a 
bending power of about \SI{4}{\tesla\metre}, and three stations of 
silicon-strip detectors and straw drift tubes placed downstream of the magnet.
The tracking system provides a measurement of the momentum of charged particles 
with a relative uncertainty that varies from \SI{0.5}{\percent} at low momentum 
to \SI{1.5}{\percent} at \SI{200}{\GeVc}.
The minimum distance of a track to a PV, the impact parameter, is measured with 
a resolution of $(15 + 29/\pt)\,\si{\micro\metre}$, where \pt is the component 
of the momentum transverse to the beam, in \si{\GeVc}.
Different types of charged hadrons are distinguished using information from two 
ring-imaging Cherenkov detectors.
Photons, electrons, and hadrons are identified by a calorimeter system 
consisting of scintillating-pad and preshower detectors, an electromagnetic 
calorimeter, and a hadronic calorimeter.
Muons are identified by a system composed of alternating layers of iron and 
multiwire proportional chambers.
To control possible left-right interaction asymmetries, the polarity of the 
dipole magnet is reversed periodically throughout data-taking.
The configuration with the magnetic field vertically upwards (downwards) bends 
positively (negatively) charged particles in the horizontal plane towards the 
centre of the LHC\@.

The online event selection is performed by a trigger, which consists of a 
hardware stage, based on information from the calorimeter and muon systems, 
followed by a two-stage software trigger, which applies first a simplified and 
then a full event reconstruction.
For the dataset used for the present analysis, at the hardware trigger stage 
the presence of a high-\pt muon candidate is required.
In the first stage of the software trigger, this candidate must be matched to a 
good-quality track which is inconsistent with originating directly from any PV 
and has a \pt above \SI{1}{\GeVc}.
The second stage requires a two-, three-, or four-track secondary vertex with a 
significant displacement from all PVs, where at least one of the tracks is 
consistent with being a muon.
At least one charged particle must have $\pt > \SI{1.6}{\GeVc}$ and must be 
inconsistent with originating from any PV\@.
A multivariate algorithm~\cite{BBDT} is used for the identification of 
secondary vertices consistent with the decays of beauty hadrons.

In the offline selection, tracks are selected on the criteria that they have a 
significant impact parameter with respect to all PVs, and also on the particle 
identification information being consistent with one of the proton, kaon, or 
pion hypotheses.
Sets of three tracks with \mbox{$\sum \pt > \SI{1.8}{\GeVc}$} are combined to 
form \Lcp candidates.
Each candidate is required to be displaced significantly from all PVs, to have 
a good quality vertex, and to have an invariant mass between 2230 and 
\SI{2350}{\MeVcc}.
The \Lcp candidate is combined with a displaced muon to form the \Lb candidate, 
which must have a good quality vertex and also satisfy the invariant mass 
requirement $2.5 < m(\Lcp\mun) < \SI{6}{\GeVcc}$.
The offline \Lb candidate is required to be matched to the candidate formed in 
the second stage of the software trigger.

Contributions from the Cabibbo-favoured decays \LcTopKz and \LcToLpi, and their 
charge conjugates, are observed in the background-subtracted $m(\pipi)$ and 
$m(\proton\pim)$ spectra.
These are removed by applying a veto in the two-pion invariant mass spectrum 
$485 < m(\pipi) < \SI{510}{\MeVcc}$ to remove \KS meson contributions, and in 
the proton-pion invariant mass spectrum $1110 < m(\proton\pim) < 
\SI{1120}{\MeVcc}$ to remove \Lz baryons.
Contributions from misidentified charm meson and background \Lcp decays are 
removed by applying a \SI{16}{\MeVcc} wide veto centred on the world average 
mass value~\cite{PDG2017} for the charm hadron in question.
Such vetoes are applied in the misidentified mass distributions for the 
following backgrounds: \DpToKKpi, \DsToKKpi, and \LcTopKpi for \pKK candidates; 
and \DpToKpipi, \DpToKKpi, \DsToKKpi, and \DsToKpipi for \ppipi candidates.

After the selection, less than \SI{2}{\percent} of the events contain more than 
one \Lb candidate.
All candidates are kept for the rest of the analysis, as other techniques of 
dealing with multiple candidates per event have been shown to be biased for 
asymmetry measurements~\cite{Koppenburg:2017zsh}.

The data were taken at two centre-of-mass energies, \sqrtseq{7} in 2011 and 
\SI{8}{\TeV} in 2012, and with two configurations of the dipole magnet 
polarity.
As the experimental efficiencies vary with these conditions, due to different 
momentum production spectra and the left-right asymmetries in the detector 
construction, the data are split into four independent subsamples by 
centre-of-mass-energy and magnet polarity.
Each stage of the analysis is carried out on each subsample independently, and 
then the individual results are combined in an average as described in 
\cref{sec:results}.

Simulated \pp collisions are used to determine experimental efficiencies and 
are generated using \pythia~\cite{Sjostrand:2007gs,*Sjostrand:2006za} with a 
specific \lhcb configuration~\cite{LHCb-PROC-2010-056}.
Particle decays are described by \evtgen~\cite{Lange:2001uf}, in which 
final-state radiation is generated using \photos~\cite{Golonka:2005pn}.
The interaction of the generated particles with the detector, and its response,
are implemented using the \geant toolkit~\cite{Allison:2006ve, 
*Agostinelli:2002hh} as described in Ref.~\cite{LHCb-PROC-2011-006}.
 
\section{Mass spectrum parameterisation}
\label{sec:mass_fit}

\begin{figure}[tb]
  \includegraphics[width=0.5\linewidth]{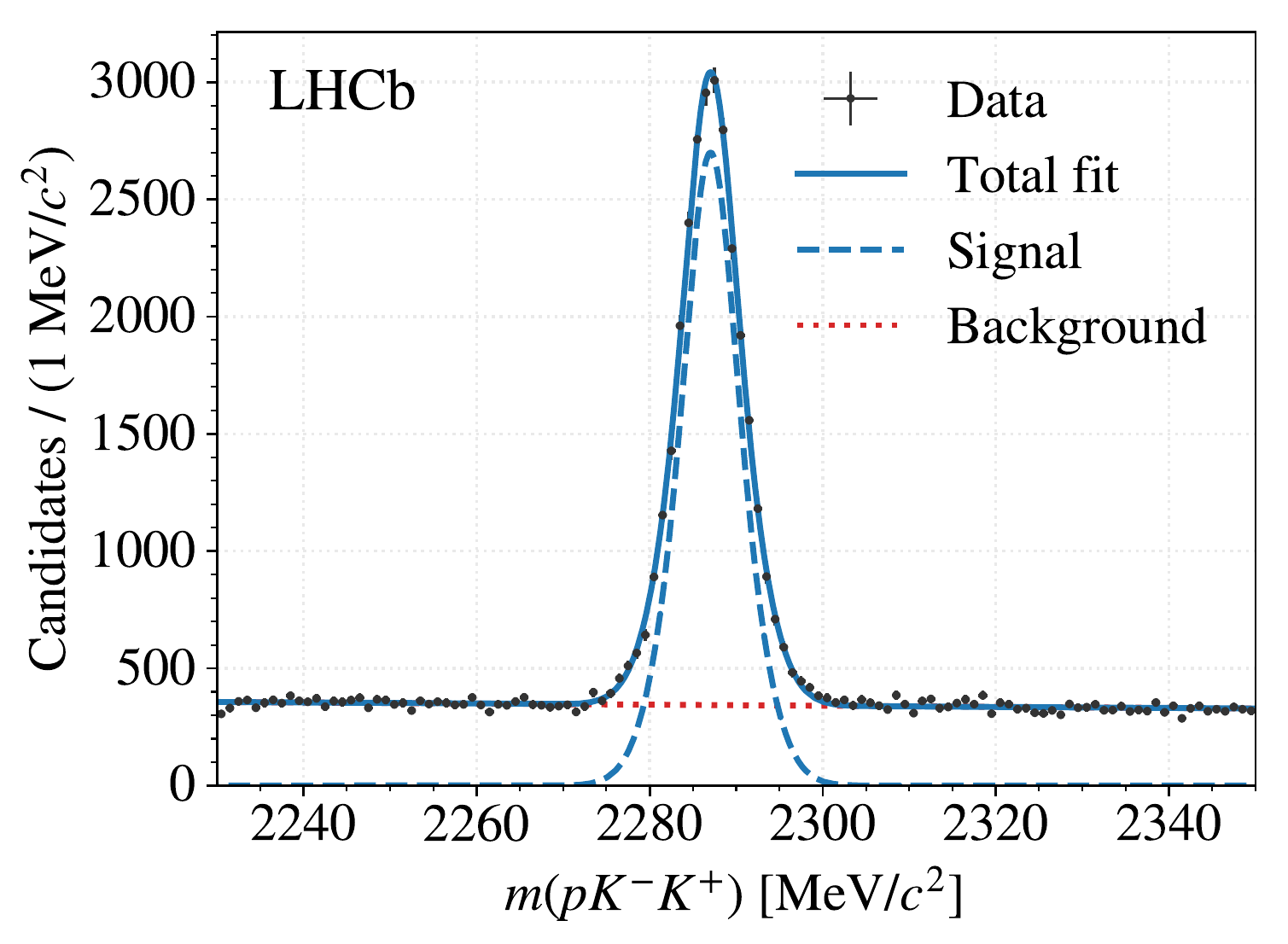}
  \includegraphics[width=0.5\linewidth]{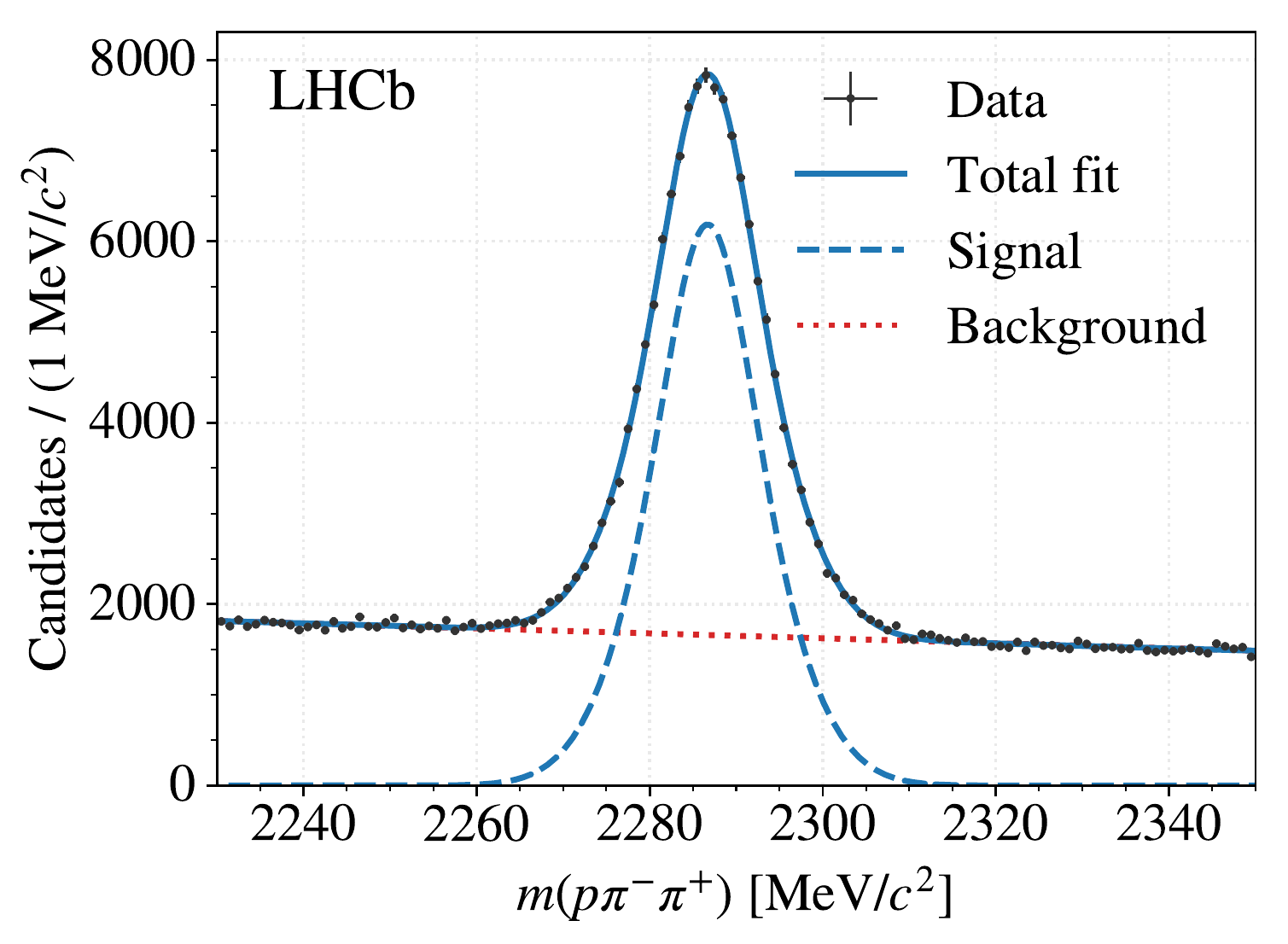}
  \caption{
    The \phh invariant mass spectra from the fully selected \LcTopKK (left) and 
    \LcToppipi (right) datasets summed over all data-taking conditions.
    The results of the fit to each dataset are shown for illustration.
    The widths of the signal distributions differ due to the different 
    $Q$-value between the two decays, where the larger value for the \LcToppipi 
    mode results in a broader shape.
  }
  \label{fig:mass_fit:fits}
\end{figure}

The \phh invariant mass is used as a discriminating variable between signal and 
combinatorial background.
Fits to the mass spectrum, shown in \cref{fig:mass_fit:fits}, are used to 
measure the \phh signal yields in order to compute \ARaw, as defined in 
\cref{eqn:formalism:araw}.
The \sPlot procedure~\cite{Pivk:2004ty} is employed to statistically subtract 
the combinatorial background component in the data, as required for the 
kinematic weighting procedure, and takes the fitted model as input.

The chosen fit model is the sum of a signal component and a background 
component, each weighted by a corresponding yield parameter.
The signal is modelled as the sum of two Gaussian distributions which share a 
common mean but have separate width parameters, and the combinatorial 
background is modelled as a first-order polynomial.

A cost function is defined as Neyman's \chisq,
\begin{equation}
  \chisq = \sum_{i = 1}^{B} \frac{(N_{i} - N\ftot(\fitvar_{i}; \paramvec))^{2}}{N_{i}},
  \label{eqn:mass_fit:objective}
\end{equation}
where $i$ is the bin index over the number of bins $B$ in the $m(\phh)$ 
spectrum, $N_{i}$ is the observed number of entries in the $i$th bin, $N$ is 
the expected number of entries in the dataset as the sum of the fitted signal 
and background yield parameters, and $\ftot(\fitvar_{i}; \paramvec)$ represents 
the integral of the total model in the $m_{i}$ bin with parameter vector 
\paramvec.
The binning is set as 120 bins of width \SI{1}{\MeVcc} in the range $2230 < 
\fitvar(\phh) < \SI{2350}{\MeVcc}$.
Fits to the \pKK and \ppipi data, summed over all conditions, are shown in 
\cref{fig:mass_fit:fits}.
A good description of the data by the model is seen in all fits to the data 
subsamples.
The \pKK and \ppipi signal yields, separated by data-taking conditions, are 
given in \cref{tab:mass_fit:yields}.

\begin{table}
  \centering
  \caption{
    Signal yields measured in the fit for each of the four subsets of the \pKK 
    and \ppipi data (two centre-of-mass energies, 7 and \SI{8}{\TeV}, and two 
    polarities of the dipole magnet, up and down).
    The corresponding integrated luminosity of each subset is also given.
  }
  \label{tab:mass_fit:yields}
  \begin{tabular}{ccSSS}
  \toprule
  \sqrts & Polarity & {Int.\ lumi.\ [pb$^{-1}$]} & {\pKK yield} & {\ppipi yield} \\
  \midrule
$7\,$TeV & Up & 422 \pm 7 & 2880 \pm 70 & 18450 \pm 190 \\
$7\,$TeV & Down & 563 \pm 9 & 3940 \pm 80 & 25130 \pm 230 \\
$8\,$TeV & Up & 1000 \pm 11 & 9040 \pm 120 & 57730 \pm 350 \\
$8\,$TeV & Down & 992 \pm 11 & 9330 \pm 120 & 60080 \pm 360 \\
  \bottomrule
\end{tabular} \end{table}

To measure \ARaw as in \cref{eqn:formalism:araw}, each data subsample is split 
by proton charge into \Lcp and \Lcm subsets.
The model used in the previously described fit is used to define 
charge-dependent models, where the parameter vectors of each model, \paramvecp 
and \paramvecm, are independent.
Rather than fitting charge-dependent signal and background yields directly, 
however, they are parameterised using the total number of signal and background 
candidates, \Nsig and \Nbkg, and the signal and background asymmetries, \ARaw 
and \ARawBkg,
\begin{align}
  \Nsigpm &= \frac{1}{2}\Nsig(1 \pm \ARaw),\\
  \Nbkgpm &= \frac{1}{2}\Nbkg(1 \pm \ARawBkg).
\end{align}
The addition of per-candidate weights, which are described in the following 
section, requires a cost function that uses the sum of weights in each bin, 
rather than the count as in \cref{eqn:mass_fit:objective}, defined as 
\begin{equation}
  \chisq = \sum_{i = 1}^{B} \left[
    \frac{(\bini^{+} - N^{+}\ftot^{+}(m_{i}; \paramvecp))^{2}}{(\binierr^{+})^{2}} +
    \frac{(\bini^{-} - N^{-}\ftot^{-}(m_{i}; \paramvecm))^{2}}{(\binierr^{-})^{2}}
  \right],
  \label{eqn:mass_fit:objective_weighted}
\end{equation}
where \binipm is the sum of the weights of candidates in the $i$th bin in the 
\Lcpm sample, and $\binierr^{\pm}$ is the uncertainty on that sum.
 
\section{Kinematic and efficiency corrections}
\label{sec:corrections}

The experimental asymmetries listed in \cref{eqn:formalism:araw_sum} are 
specific to the production environment at the \lhc and the construction of the 
\lhcb detector, and so their cancellation in \dACPwgt is crucial in providing 
an unbiased measurement.
This section presents the statistical methods used to compute the kinematic and 
efficiency corrections, which are evaluated as per-candidate weights to be used 
in the simultaneous \chisq fit previously described.

\subsection{Kinematic weighting}
\label{sec:corrections:kinematics}

The production and detection asymmetries depend on the kinematics of the 
particles involved.
If the \Lb, muon, and proton kinematic spectra are the same between the \pKK 
and \ppipi data, then the \Lb production asymmetry and muon and proton 
detection asymmetries will cancel in \dACPwgt.
If the \hm and \hp kinematics are equal \emph{within} each separate \pKK and 
\ppipi sample, then the kaon ($f = \pKK$) or pion ($f = \ppipi$) detection 
asymmetries will cancel in $\ARaw(f)$.
The \hm kinematics agree well with the \hp spectra in the data, but the \Lb, 
muon, and proton kinematics do not, and so a per-candidate weighting technique 
is employed to match the kinematic spectra of the $\ppipi\mun$ state to those 
of the $\pKK\mun$ state.

To compute the per-candidate weights, a forest of shallow decision trees with 
gradient boosting (a GBDT) is 
used~\cite{Friedman00,Friedman01,Rogozhnikov:2016bdp}.
This method recursively bins the \pKK and \ppipi input data such that regions 
with larger differences between the two samples are more finely partitioned.
After fitting, each \ppipi candidate is assigned a weight $d$.
To reduce biases that may result from overfitting, where the GBDT model becomes 
sensitive to the statistical fluctuations in the input data, the data are split 
in two, and independent GBDTs are fitted to each subset.
The GBDT built with one half of the data is used to evaluate weights for the 
other half, and vice versa.

The \Lcp and proton \pt and pseudorapidity for each \pKK and \ppipi candidate 
are used as input to the GBDT.
The \Lcp kinematics are chosen since the large boost in the laboratory frame 
induces a large correlation with \Lb and muon kinematics.
An agreement in the \Lcp kinematics therefore results in an agreement in the 
\Lb and muon spectra.
The proton kinematics are chosen as the different $Q$ values of the decays will 
a priori result in different proton spectra.

\begin{figure}[tb]
  \includegraphics[width=0.5\linewidth]{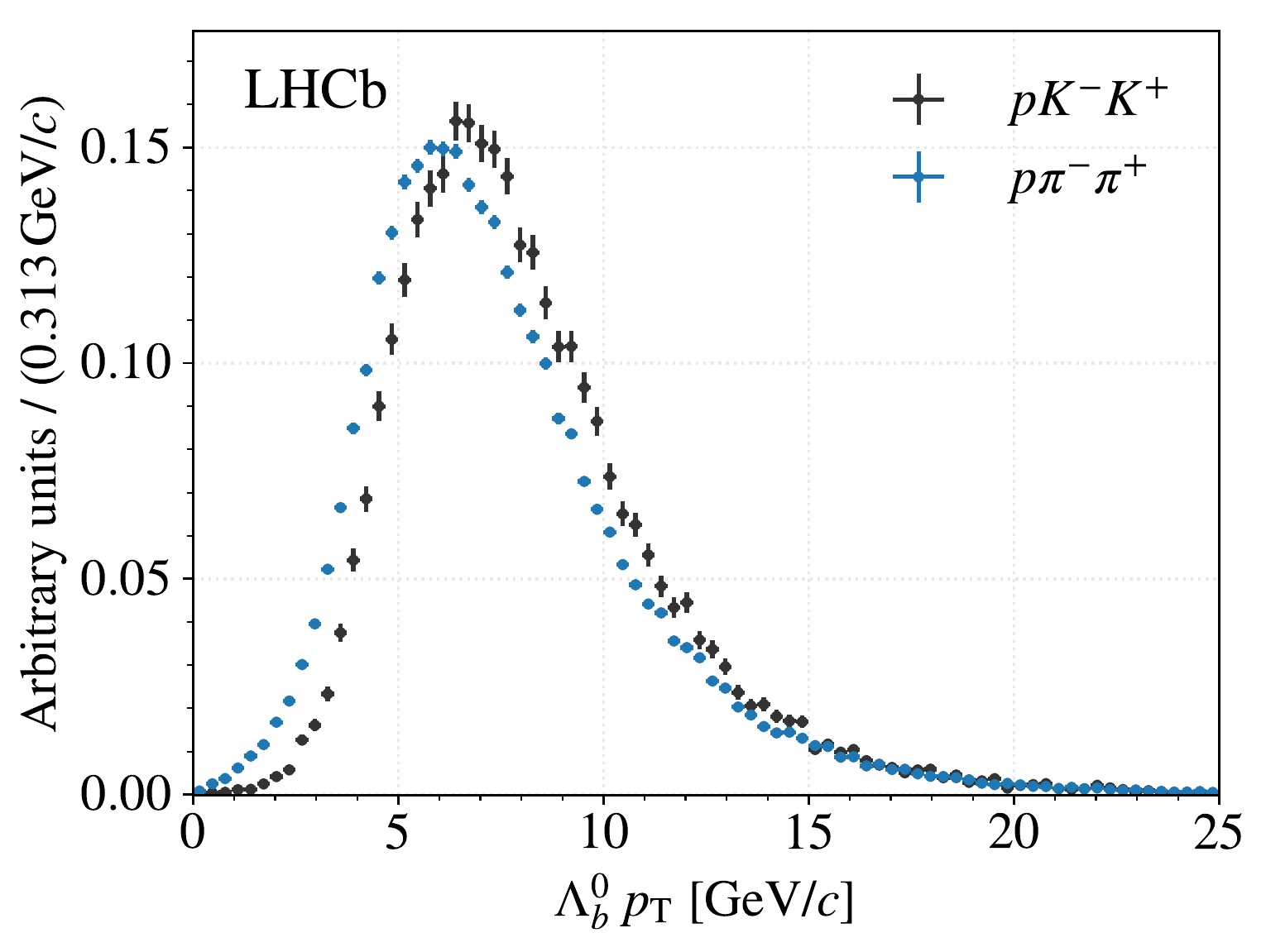}
  \includegraphics[width=0.5\linewidth]{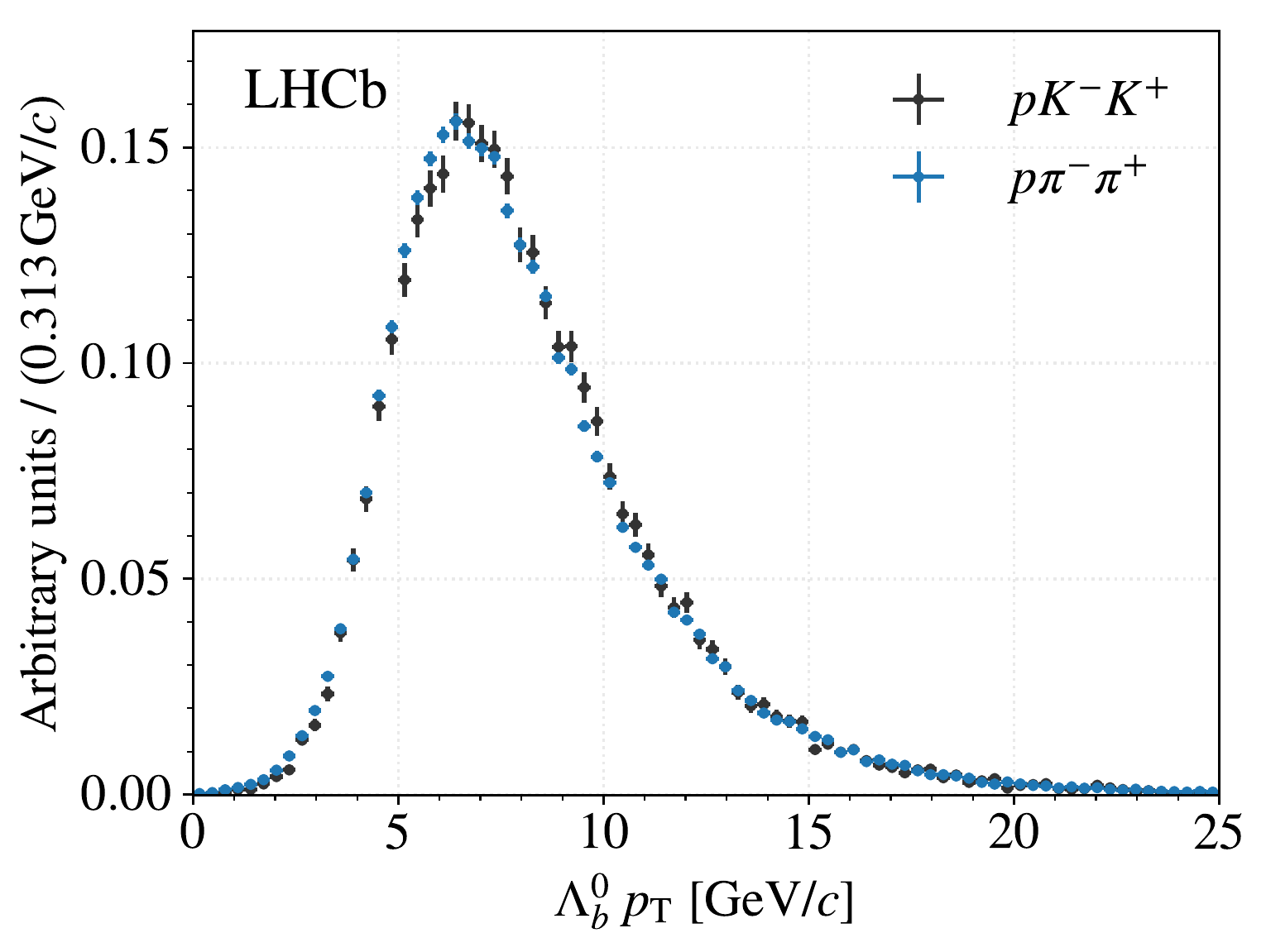}

  \includegraphics[width=0.5\linewidth]{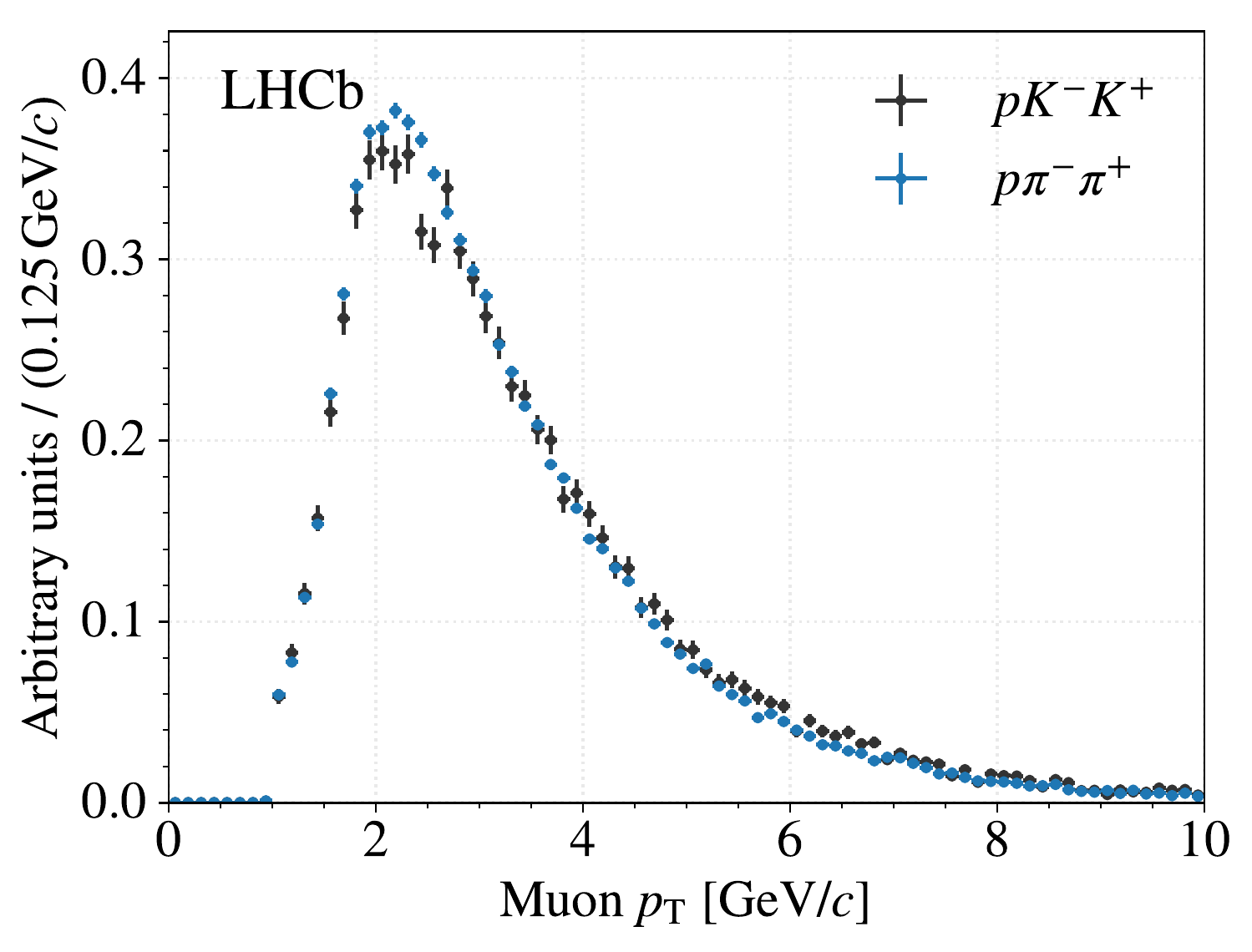}
  \includegraphics[width=0.5\linewidth]{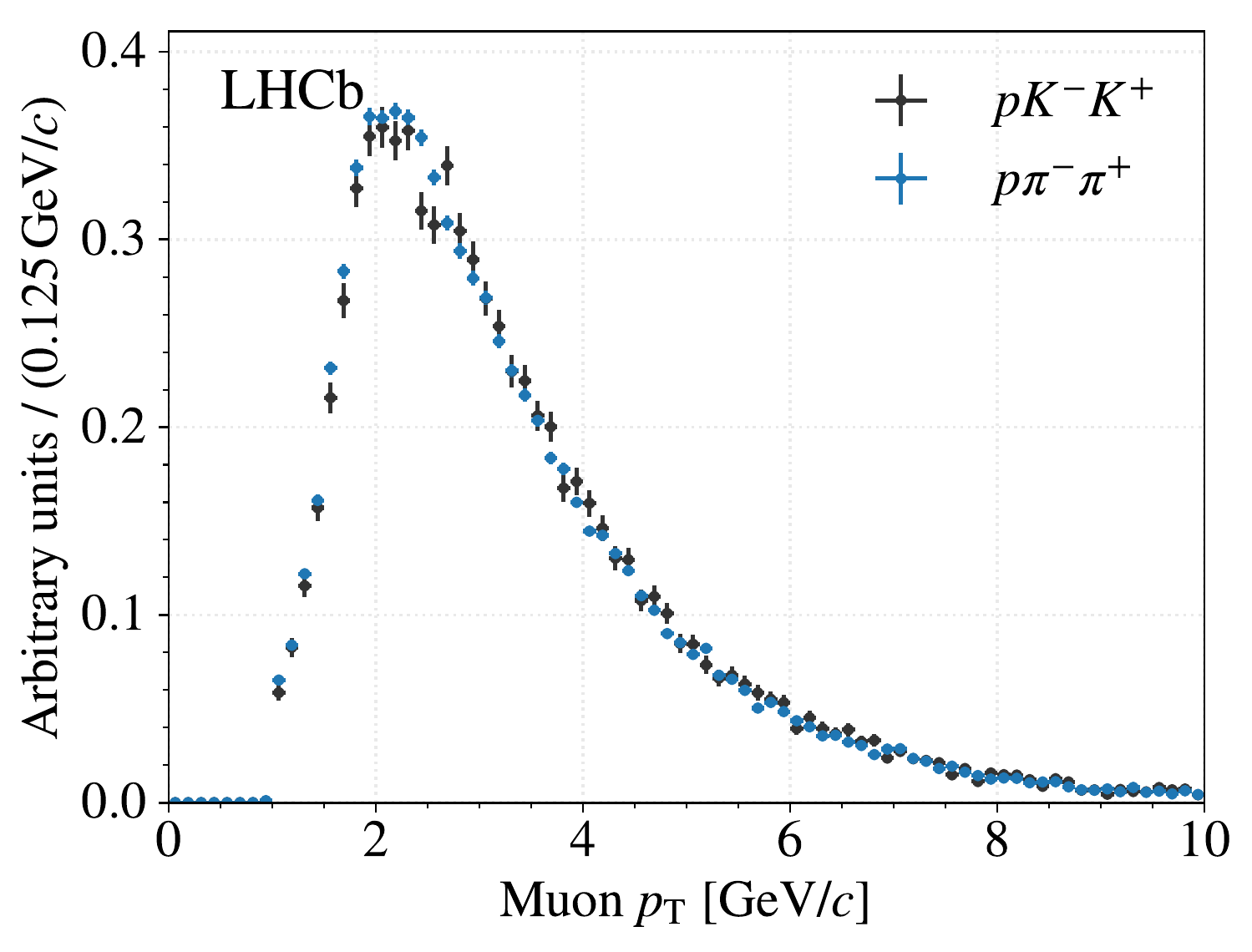}

  \includegraphics[width=0.5\linewidth]{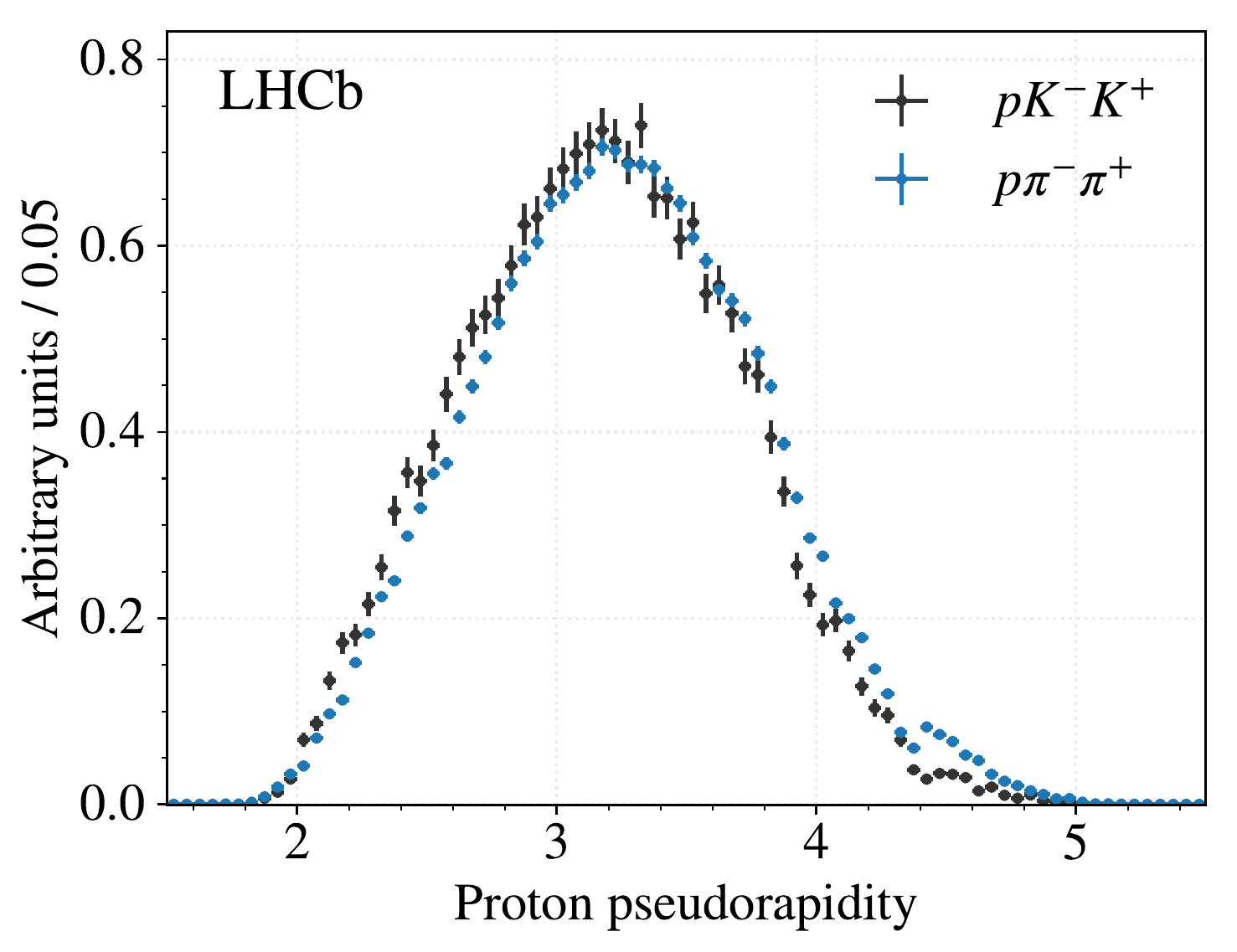}
  \includegraphics[width=0.5\linewidth]{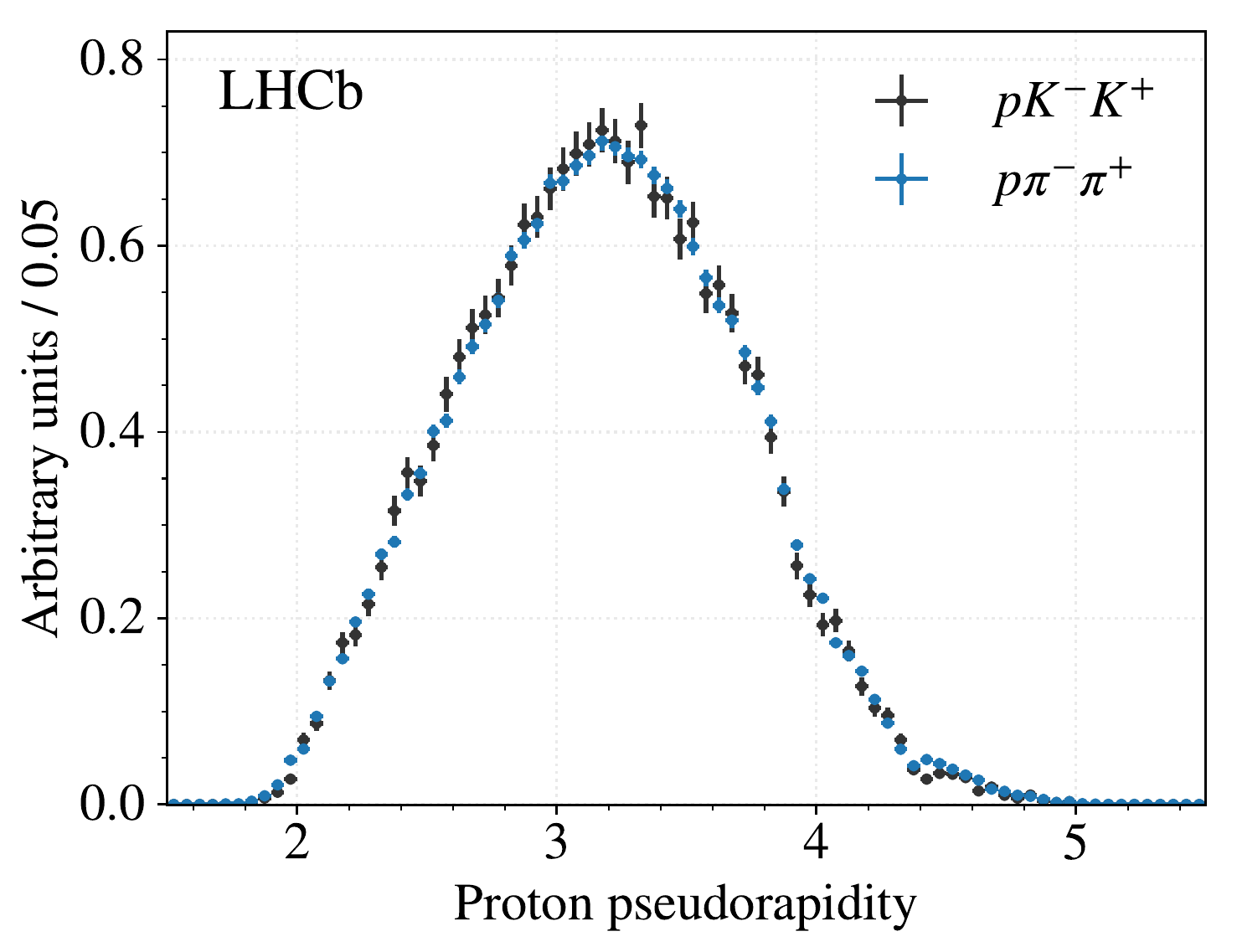}
  \caption{
    Background-subtracted distributions of the \Lb candidate transverse 
    momentum (top row), the muon candidate transverse momentum (middle row), 
    and the proton candidate pseudorapidity (bottom row) both before (left 
    column) and after (right column) weighting the \ppipi sample (blue points) 
    to match the \pKK sample (black points).
    The data are summed across all data-taking conditions.
  }
  \label{fig:correction:kinematic_weighting}
\end{figure}

The \Lb, muon, proton, and \hm/\hp kinematics agree well after weighting, as 
demonstrated for a subset of kinematic variables in 
\cref{fig:correction:kinematic_weighting}.
Any remaining differences will result in residual asymmetries in \dACPwgt, and 
the presence of these differences is studied in the context of systematic 
effects as described in \cref{sec:systematics}.

\subsection{Efficiency corrections}
\label{sec:corrections:efficiency}

The acceptance, reconstruction, and selection efficiencies as a function of the 
5D \phh phase space are also modelled using GBDTs.
Simulated events are generated with a uniform \LcTophh matrix element and used 
as input to the training, sampled before and after the detector acceptance and 
data processing steps.
One- and two-dimensional efficiency estimates are made as histogram ratios of 
the before and after data, and projections of the efficiency model obtained 
using the simulation agrees well with these.
The model is then used to predict per-candidate efficiencies in the data.

\subsection{Use in determining \texorpdfstring{\boldmath{\ARaw}}{ARaw}}
\label{sec:corrections:usage}

\begin{figure}[tb]
  \includegraphics[width=0.5\linewidth]{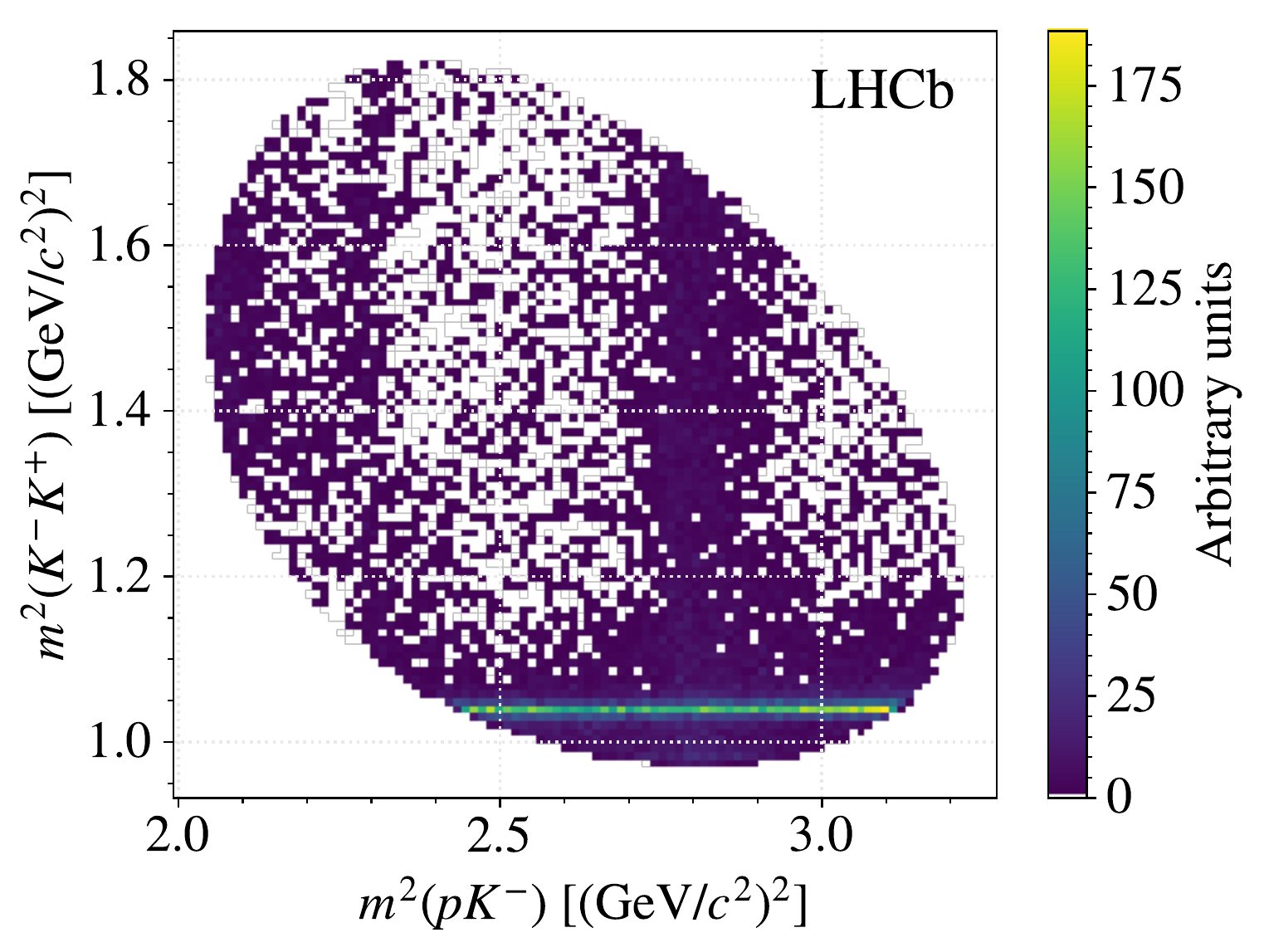}
  \includegraphics[width=0.5\linewidth]{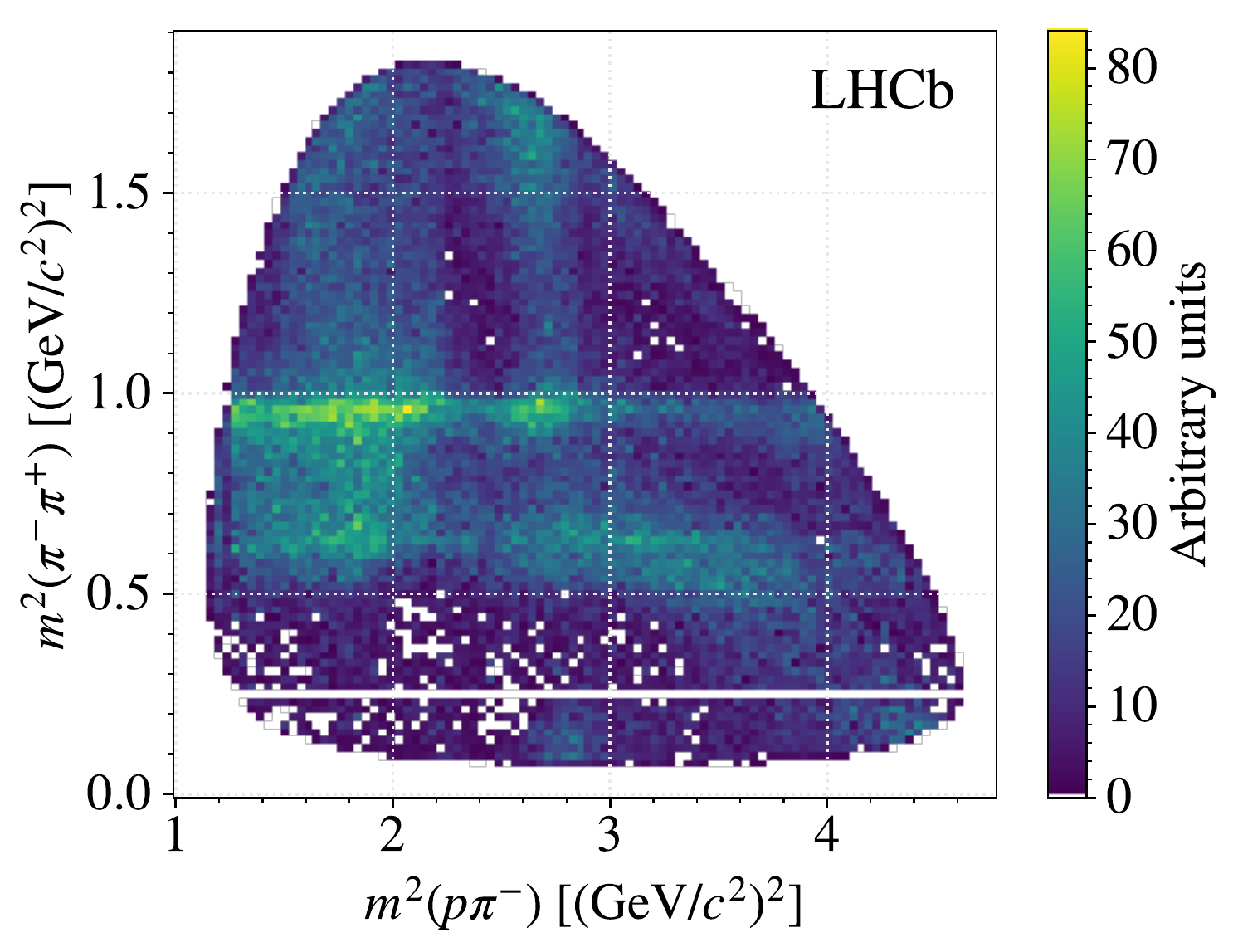}
  \caption{
    Background-subtracted and efficiency-corrected \LcTopKK (left) and 
    \LcToppipi (right) data in the \msqphm--\msqhh plane, integrated across all 
    data-taking subsamples.
    The \pKK data feature a prominent \phiToKK component, whilst the \ppipi 
    data exhibit $\decay{\rho(770)/\omega(782)}{\pipi}$ and 
    $\decay{f_{0}(980)}{\pipi}$ components.
  }
  \label{fig:correction:phase_space}
\end{figure}

The cost function in \cref{eqn:mass_fit:objective_weighted} uses the sum of 
per-candidate weights in each bin and its uncertainty.
The weights are defined using the kinematic weight $d_{j}$, equal to unity for 
\LcTopKK candidates, and the efficiency correction $\eff_{j}$ of the $j$th 
candidate in the $i$th $m(\phh)$ bin,
\begin{equation}
  \bini = W\sum_{j = 1}^{N_{i}} \frac{
    d_{j}
  }{
    \eff_{j}
  },
  \quad
  \binierr^{2} = \bini,
\end{equation}
where the normalisation factor $W$ is defined as
\begin{equation}
  W = \frac{
    \sum_{j = 1}^{N_{i}} \frac{d_{j}}{\eff_{j}}
  }{
  \sum_{j = 1}^{N_{i}} \left(\frac{d_{j}}{\eff_{j}}\right)^{2}
  }.
\end{equation}
The term $W_{i}$ can be called the number of `effective' entries in the bin, as 
it encodes the size of an unweighted data sample with the same statistical 
power as the weighted sample.
For the \ppipi data, which is weighted to match the \pKK kinematics, the 
effective sample size is around \SI{80}{\percent} that of the unweighted \ppipi 
sample.
The weighted data are shown in the \msqphm--\msqhh plane in 
\cref{fig:correction:phase_space}.

The statistical treatment of the weights in the fit is validated by randomly 
sorting candidates into \Lcp and \Lcm datasets and fitting the model 500 times, 
where it is seen that the distribution of $\ARaw(f)$ divided by its uncertainty 
is centred around zero, the expected value, and has a standard deviation of 1, 
indicating that the error estimate is correct.

\section{Systematic effects}
\label{sec:systematics}

To evaluate possible biases on the measurement of \dACPwgt due to systematic 
effects, several studies are performed and deviations from the nominal results 
are computed.
Statistically significant deviations are assigned to the measurement as 
systematic uncertainties.

The model used in the simultaneous \chisq fit, described in 
\cref{sec:mass_fit}, is derived empirically, and there may be other models 
which described the data similarly well.
Variations of the choice of background model are found to have a negligible 
effect on the measurement of \dACPwgt, however different signal models can 
change the results significantly.
To assess an associated systematic uncertainty based on the choice of signal 
model, the signal \Lcp and \Lcm yields are determined using the method of 
sideband subtraction.
Here, data from the regions on either side of the \Lcpm signal peak are assumed 
to be linearly distributed and are used to approximate the background yield in 
the peak region.
Given that the data used for sideband subtraction are the same as for the 
nominal \chisq fit, the measurements using the two techniques are assumed to be 
fully correlated, such that even small differences between them are 
statistically significant.
On the average of \dACPwgt, taken across all data-taking conditions, a 
difference of \SI{0.2}{\percent} is seen with respect to the average of the 
results using the full fit, and this difference is assigned as a systematic 
uncertainty.

The kinematic weighting procedure defined in \cref{sec:corrections} can only 
equalise the $\pKK\mun$ and $\ppipi\mun$ kinematics approximately, and so 
residual differences will remain.
These differences can cause a bias on \dACPwgt, with a size depending on the 
size of the relevant asymmetry.
Measurements of the \Lb production asymmetry and the muon, kaon, and pion 
detection asymmetries using LHCb data 
exist~\cite{LHCb-PAPER-2015-032,LHCb-PAPER-2016-013,LHCb-PAPER-2014-013,LHCb-PAPER-2012-009}, 
and estimates of the proton detection asymmetry using simulated events have 
been used previously~\cite{LHCb-PAPER-2015-032},
and so the measurement of \dACPwgt can be corrected for directly.
The correction is found to be less than one per mille, but with a relative 
uncertainty of \SI{20}{\percent}, and so a systematic uncertainty of 
\SI{0.1}{\percent} is assigned to \dACPwgt.

The limited size of the simulated sample results in a statistical uncertainty 
on the efficiencies taken from the phase space efficiency model.
The size of this uncertainty is evaluated by resampling the simulated data 500 
times, each time building a new model and computing the efficiencies of the 
data using that model.
The simultaneous \chisq fit to measure $\ARaw(f)$ is then performed for each 
set of efficiencies, resulting in a spread of values of \dACPwgt with a 
standard deviation of \SI{0.57}{\percent}, which is taken as the systematic 
uncertainty due to the limited simulated sample size.

Due to the presence of \Lcp decays originating from sources other than 
\LbToLcmuX decays, such as directly from the PV or from other \bquark-hadron 
decays, the measurement may be biased, as such sources can carry different 
experimental asymmetries.
The composition of the data sample is inferred from the reconstructed \Lb mass 
and from the impact parameter distribution of the \Lcp vertex.
The latter is seen to be consistent with that for \Lcp produced exclusively in 
\bquark-hadron decays, whilst the former is consistent between \pKK and \ppipi 
samples, such that asymmetries from other sources will cancel in \dACPwgt.
Any associated systematic uncertainty is assumed to be negligible.

The total systematic uncertainty is found to be \SI{0.61}{\percent}, computed 
as the sum in quadrature of the individual uncertainties. These are assumed to 
be uncorrelated and are summarised in Table 2.

\begin{table}
  \centering
  \caption{
    Systematic uncertainties on \dACPwgt and their magnitudes.
    The dash indicates that the uncertainty is assessed to be negligible.
  }
  \label{tab:systematics:summary}
  \begin{adjustbox}{center}
    \begin{tabular}{rS[table-format=1.2]}
  \toprule
  Source & {Uncertainty [\si{\percent}]} \\
  \midrule
  Fit signal model              & 0.20 \\
  Fit background model          & {\textemdash} \\
  Residual asymmetries          & 0.10 \\
  Limited simulated sample size & 0.57 \\
  Prompt \Lcp{}                 & {\textemdash} \\
  \midrule
  Total                         & 0.61 \\
  \bottomrule
\end{tabular}
   \end{adjustbox}
\end{table}
 
\section{Results}
\label{sec:results}

\begin{figure}
  \includegraphics[width=0.5\textwidth]{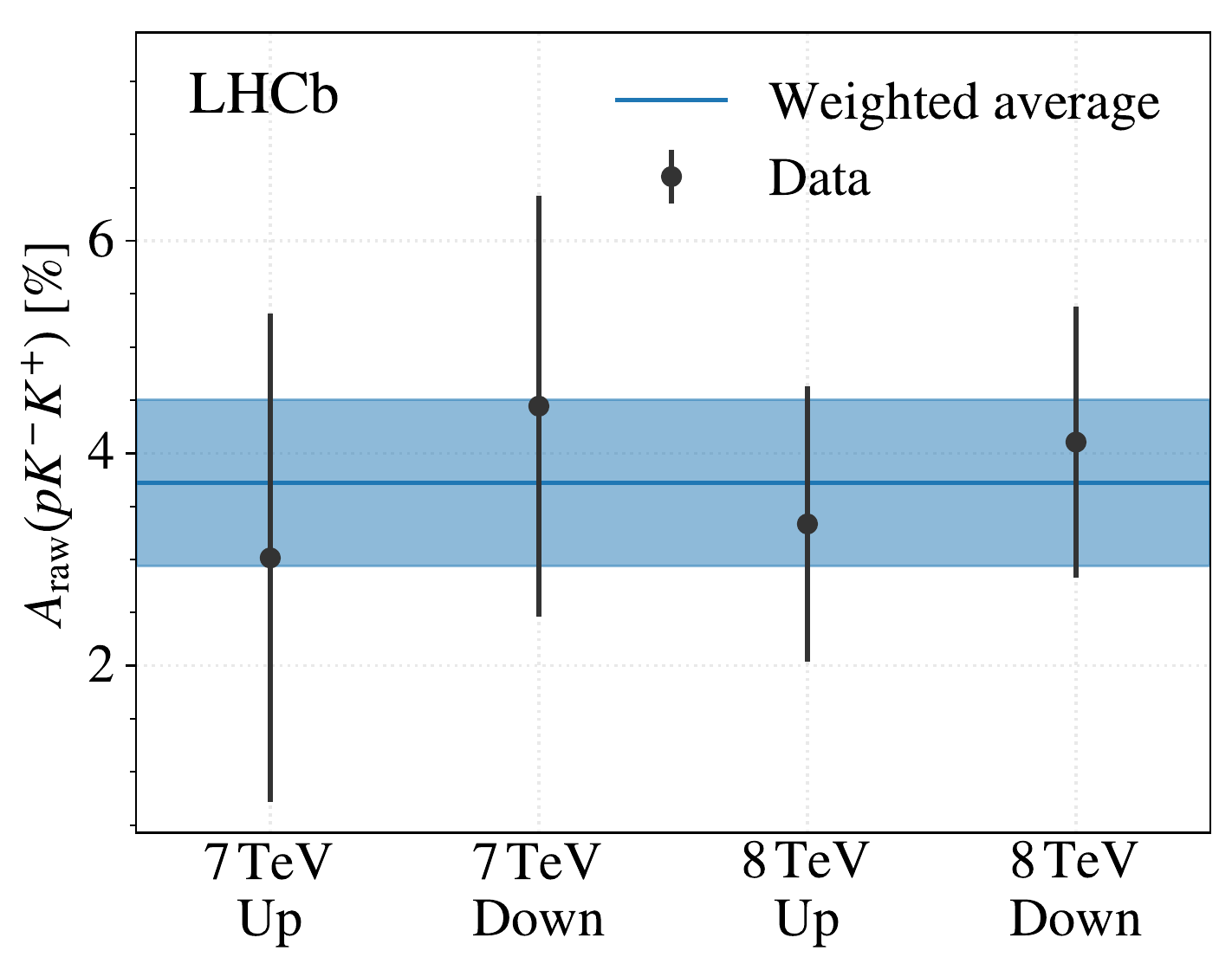}
  \includegraphics[width=0.5\textwidth]{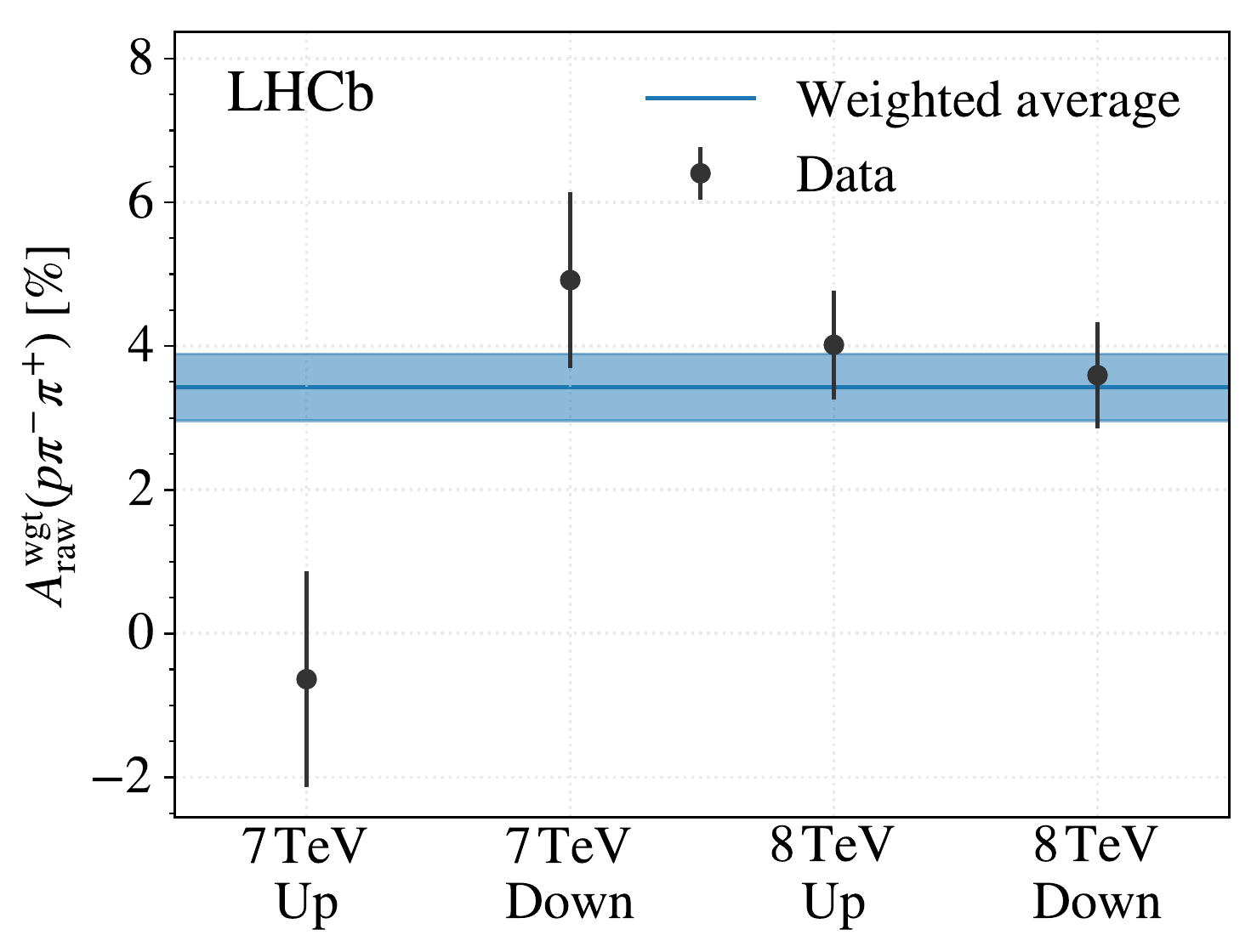}
  \begin{center}
    \includegraphics[width=0.5\textwidth]{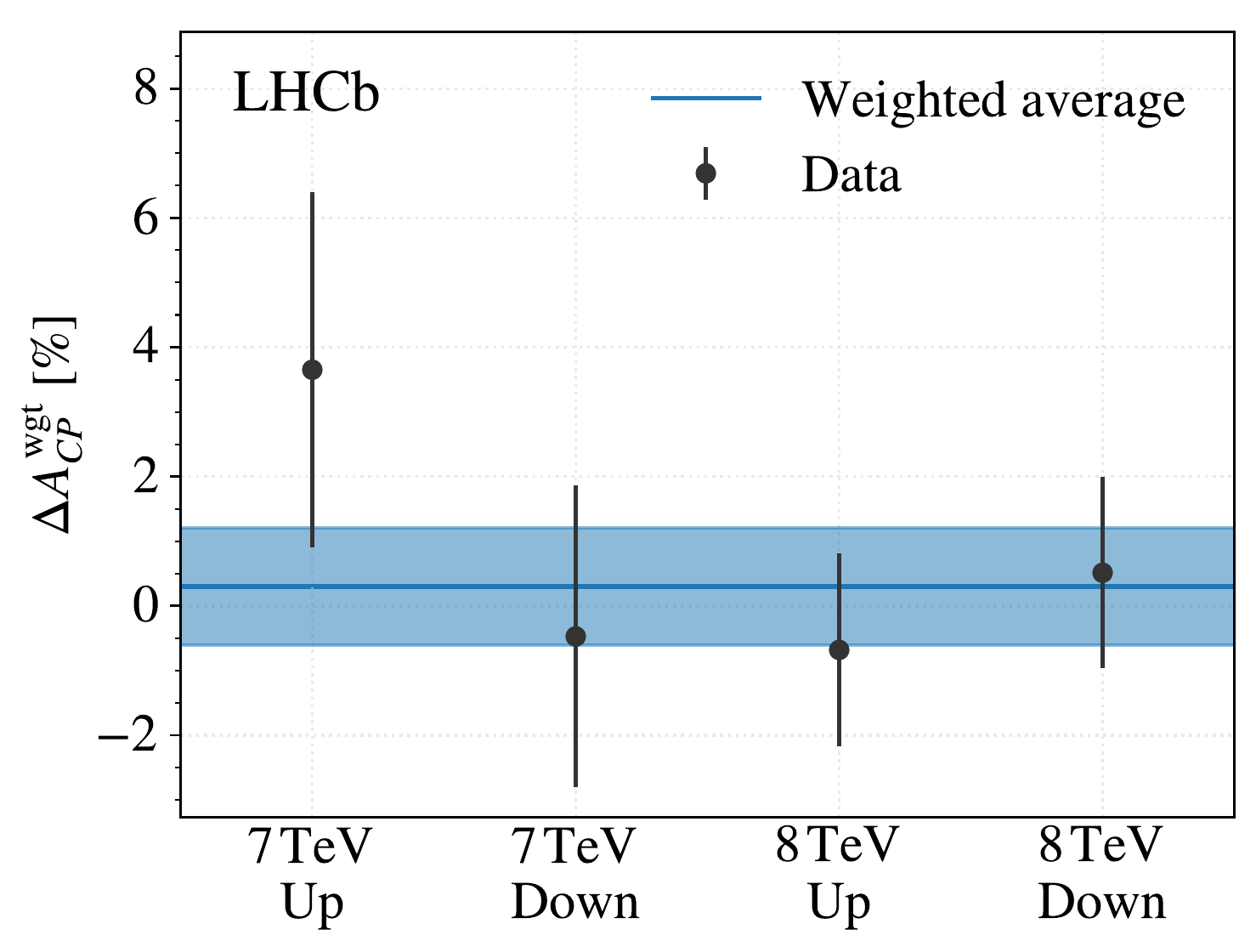}
  \end{center}
  \caption{
    Values of and statistical uncertainties on the asymmetries $\ARaw(\pKK)$ 
    (top left), $\ARawwgt(\ppipi)$ (top right), and \dACPwgt (bottom centre), 
    for the four data subsamples (two centre-of-mass energies, 7 and 
    \SI{8}{\TeV}, and two polarities of the dipole magnet, up and down).
    For each asymmetry, the average of the four data points, as described in 
    the text, is also shown, where the band indicates the uncertainty.
  }
  \label{fig:results:asymmetries}
\end{figure}

The value of $\ARaw(f)$ is found for each final state and data-taking condition 
separately, and for a given centre-of-mass energy is taken as the arithematic 
average of the polarity-dependent measurements.
The average across \sqrtseq{7} and $\SI{8}{\TeV}$ is made by weighting the 
measurements by their variances.
The asymmetries for \pKK and \ppipi are measured to be
\begin{align*}
  \ARaw(\pKK) &= \ARawpKK,\\
  \ARawwgt(\ppipi) &= \ARawppipi.
  \label{eqn:results:araw}
\end{align*}
where the uncertainties are statistical and take into account the reduction in 
statistical power due to the weighting.
The difference is
\begin{equation*}
  \dACPwgt = (\deltaACPval \pm \deltaACPunc \pm 0.61)\si{\deltaACPunit},
  \label{eqn:results:dacp}
\end{equation*}
where the first uncertainty is statistical and the second is systematic.
The measurements of $\ARaw(\pKK)$, $\ARawwgt(\ppipi)$, and \dACPwgt as a 
function of data-taking conditions are presented in 
\cref{fig:results:asymmetries}.

\section{Summary}
\label{sec:summary}

The raw \CP asymmetries in the decays \LcTopKK and \ppipi are measured using 
\LbToLcmuX decays.
Kinematics in the \ppipi data are weighted to match those in the \pKK data, 
such that the effect of experimental asymmetries on the \CP asymmetry parameter 
\dACPwgt is negligible.
Acceptance, reconstruction, and selection efficiencies across the 
five-dimensional \phh phase space are corrected for.
Systematic effects arising from the mass distribution modelling, imperfect 
kinematic weighting, finite simulated sample size, and the inclusion of \Lcp 
decays from sources other than \LbToLcmuX decays are considered.
The total systematic uncertainty assigned to these effects is smaller than the 
statistical uncertainty on \dACPwgt, whose central value is measured to be 
consistent with zero.

This analysis constitutes the first measurement of a \CP violation parameter in 
three-body \Lcp decays, but more data is required to match the sensitivity of 
similar measurements using charm mesons.
Further studies into the structure of the \phh phase space, across which 
\CP-violating effects may strongly vary, would be beneficial as input to 
theoretical calculations.
 
\section*{Acknowledgements}
\noindent We express our gratitude to our colleagues in the CERN
accelerator departments for the excellent performance of the LHC. We
thank the technical and administrative staff at the LHCb
institutes. We acknowledge support from CERN and from the national
agencies: CAPES, CNPq, FAPERJ and FINEP (Brazil); MOST and NSFC
(China); CNRS/IN2P3 (France); BMBF, DFG and MPG (Germany); INFN
(Italy); NWO (The Netherlands); MNiSW and NCN (Poland); MEN/IFA
(Romania); MinES and FASO (Russia); MinECo (Spain); SNSF and SER
(Switzerland); NASU (Ukraine); STFC (United Kingdom); NSF (USA).  We
acknowledge the computing resources that are provided by CERN, IN2P3
(France), KIT and DESY (Germany), INFN (Italy), SURF (The
Netherlands), PIC (Spain), GridPP (United Kingdom), RRCKI and Yandex
LLC (Russia), CSCS (Switzerland), IFIN-HH (Romania), CBPF (Brazil),
PL-GRID (Poland) and OSC (USA). We are indebted to the communities
behind the multiple open-source software packages on which we depend.
Individual groups or members have received support from AvH Foundation
(Germany), EPLANET, Marie Sk\l{}odowska-Curie Actions and ERC
(European Union), ANR, Labex P2IO and OCEVU, and R\'{e}gion
Auvergne-Rh\^{o}ne-Alpes (France), RFBR, RSF and Yandex LLC (Russia),
GVA, XuntaGal and GENCAT (Spain), Herchel Smith Fund, the Royal
Society, the English-Speaking Union and the Leverhulme Trust (United
Kingdom).

\addcontentsline{toc}{section}{References}
\setboolean{inbibliography}{true}
\bibliographystyle{LHCb}
\bibliography{main,LHCb-PAPER,LHCb-CONF,LHCb-DP,LHCb-TDR}

\newpage

\ifthenelse{\boolean{paperconf}}{
 
\newpage
\centerline{\large\bf LHCb collaboration}
\begin{flushleft}
\small
R.~Aaij$^{40}$,
B.~Adeva$^{39}$,
M.~Adinolfi$^{48}$,
Z.~Ajaltouni$^{5}$,
S.~Akar$^{59}$,
J.~Albrecht$^{10}$,
F.~Alessio$^{40}$,
M.~Alexander$^{53}$,
A.~Alfonso~Albero$^{38}$,
S.~Ali$^{43}$
% ...

{\footnotesize \it
% ...
$ ^{5}$Clermont Universit{\'e}, Universit{\'e} Blaise Pascal, CNRS/IN2P3, LPC, Clermont-Ferrand, France\\
% ...
$ ^{10}$Fakult{\"a}t Physik, Technische Universit{\"a}t Dortmund, Dortmund, Germany\\
% ...
$ ^{38}$ICCUB, Universitat de Barcelona, Barcelona, Spain\\
$ ^{39}$Instituto Galego de F{\'\i}sica de Altas Enerx{\'\i}as (IGFAE), Universidade de Santiago de Compostela, Santiago de Compostela, Spain\\
$ ^{40}$European Organization for Nuclear Research (CERN), Geneva, Switzerland\\
% ...
$ ^{43}$Nikhef National Institute for Subatomic Physics, Amsterdam, The Netherlands\\
% ...
$ ^{48}$H.H. Wills Physics Laboratory, University of Bristol, Bristol, United Kingdom\\
% ...
$ ^{53}$School of Physics and Astronomy, University of Glasgow, Glasgow, United Kingdom\\
% ...
$ ^{59}$University of Cincinnati, Cincinnati, OH, United States\\
% ...
}
\end{flushleft} 
}{}

\end{document}